\pdfoutput=1
\documentclass[a4paper,UKenglish,cleveref, autoref, thm-restate]{lipics-v2021}

\title{Making Three Out of Two: Three-Way Online Correlated Selection} 

\author{Yongho Shin}{Department of Computer Science, Yonsei University, South Korea}{yshin@yonsei.ac.kr}{}{}

\author{Hyung-Chan An\footnote{Corresponding author. Department of Computer Science, Yonsei University, 50 Yonsei-ro, Seodaemun-gu,
Seoul 03722, South Korea.}}{Department of Computer Science, Yonsei University, South Korea}{hyung-chan.an@yonsei.ac.kr}{}{}

\authorrunning{Y. Shin and H.-C. An} 

\Copyright{Yongho Shin and Hyung-Chan An} 

\ccsdesc[100]{Theory of computation~Online algorithms} 

\keywords{online correlated selection, multi-way OCS, online algorithms, negative correlation, edge-weighted online bipartite matching} 

\category{} 

\relatedversion{} 

\funding{This work was supported by the National Research Foundation of Korea (NRF) grant funded by the Korea government (MSIT) (No. NRF-2019R1C1C1008934).}

\nolinenumbers 

\hideLIPIcs  

\EventEditors{John Q. Open and Joan R. Access}
\EventNoEds{2}
\EventLongTitle{42nd Conference on Very Important Topics (CVIT 2016)}
\EventShortTitle{CVIT 2016}
\EventAcronym{CVIT}
\EventYear{2016}
\EventDate{December 24--27, 2016}
\EventLocation{Little Whinging, United Kingdom}
\EventLogo{}
\SeriesVolume{42}
\ArticleNo{23}

\newcommand{\A}{\mathcal{A}}
\newcommand{\B}{\mathcal{B}}

\newcommand{\R}{\mathbb{R}}
\newcommand{\Z}{\mathbb{Z}}

\newcommand{\nextpair}[1]{\mathsf{next}(#1)}
\newcommand{\kmax}{k_\mathsf{\max}}
\newcommand{\ellmax}{\ell_\mathsf{\max}}

\newcommand{\floor}[1]{{\lfloor #1 \rfloor}}

\newcommand{\gammaA}{\gamma_\mathcal{A}}
\newcommand{\gammaB}{\gamma_\mathcal{B}}

\newcommand{\kvect}{\mathbf{k}}
\newcommand{\xvect}{\mathbf{x}}
\newcommand{\yvect}{\mathbf{y}}
\newcommand{\binomprob}{\mathsf{binom}}

\newcommand{\pred}[1]{\mathsf{pred}(#1)}

\newcommand{\jhat}{\widehat{j}}
\newcommand{\jbar}{\overline{j}}

\newcommand{\sigRtwo}{\sigma_{R_2}}
\newcommand{\sigD}{\sigma_D}

\newcommand{\delRthreeBeta}[2]{B^{R_3}_{#1, #2}}
\newcommand{\delRtwoBeta}[2]{B^{R_2}_{#1, #2}}
\newcommand{\delDBeta}[2]{B^{D}_{#1, #2}}

\newcommand{\ubar}{\overline{u}}
\newcommand{\ustar}{u^*}

\newcommand{\betaRthree}{\beta^{R_3}}
\newcommand{\betaRtwo}{\beta^{R_2}}
\newcommand{\betaD}{\beta^{D}}

\newcommand{\ybar}{\overline{y}}

\newcommand{\anhc}[1]{{#1}}
\DeclareMathOperator*{\E}{E}

\begin{document}

\maketitle

\begin{abstract}
\emph{Two-way online correlated selection (two-way OCS)} is an online algorithm that, at each timestep, takes a pair of elements from the ground set and irrevocably chooses one of the two elements, while ensuring negative correlation in the algorithm's choices. Whilst OCS was initially invented by Fahrbach, Huang, Tao, and Zadimoghaddam to solve the edge-weighted online bipartite matching problem, it is an interesting technique on its own due to its capability of introducing a powerful algorithmic tool, namely negative correlation, to online algorithms. As such, Fahrbach et al. posed two tantalizing open questions in their paper, one of which was the following: Can we obtain \mbox{\emph{$n$-way OCS}} for $n>2$, in which the algorithm can be given $n>2$ elements to choose from at each timestep?

In this paper, we affirmatively answer this open question by presenting a \emph{three-way OCS}. Our algorithm uses two-way OCS as its building block and is simple to describe; however, as it internally runs two instances of two-way OCS, one of which is fed with the output of the other, the final output probability distribution becomes highly elusive. We tackle this difficulty by approximating the output distribution of OCS by a flat, less correlated function and using it as a safe ``surrogate'' of the real distribution. Our three-way OCS also yields a 0.5093-competitive algorithm for edge-weighted online matching, demonstrating its usefulness.

\end{abstract}

\section{Introduction} \label{sect:intro}

\emph{Online correlated selection (OCS)} is an online algorithm that, at each timestep, takes a subset of the ground set as the input and irrevocably chooses a single element from the subset. When every input subset has cardinality $n$, we call it \emph{$n$-way OCS} in particular. The aim of online correlated selection is to ensure a certain level of negative correlation in the choice made by the algorithm. For example, suppose we run a two-way OCS and afterwards specify $m$ timesteps that contained some common element. The probability that this element was never chosen would be $(1/2)^m$ if the algorithm made independent and uniformly random choices; the goal of two-way OCS is to reduce this probability by introducing negative correlations. (See Definition~\ref{defn:gammaocs} for a full definition that quantifies the desired amount of reduction.)

OCS was first invented by Fahrbach, Huang, Tao, and Zadimoghaddam~\cite{fahrbach} to solve the edge-weighted online bipartite matching problem. Negative correlation has proven to be a very powerful technique in approximation algorithms design (see, e.g.,~\cite{srinivasan01distributions, asadpour2017log, chekuri09randomized} for a limited list of examples); this suggests that OCS as well bears high potential as a general tool in online algorithms design rather than as a specialized technique to solve a particular problem. This opportunity was also observed by the breakthrough paper of Fahrbach et al.~\cite{fahrbach} and recently exemplified by Huang, Zhang, and Zhang~\cite{huang2020adwords}, who devised a certain variant of OCS called \emph{panoramic OCS} to solve the AdWords problem with general bids.

In light of such value of OCS as an algorithmic tool, Fahrbach et al.~\cite{fahrbach} raised in their paper two follow-up questions that arise quite naturally: Can we improve the performance of their two-way OCS? Can we obtain an $n$-way OCS for $n>2$? This paper affirmatively answers the latter question. In this paper, we present a simple \emph{three-way OCS} and analyze its performance. 
We will also show that our three-way OCS can be used to improve the previous competitive ratio of 0.5086 due to Fahrbach et al. to give a new 0.5093-competitive algorithm for edge-weighted online bipartite matching.

In fact, the construction itself of our three-way OCS is easy to describe. It internally executes two instances of two-way OCS.
Upon arrival of a triple, we choose two of the three elements uniformly at random, and let the first two-way OCS choose one of them. We then pass its output, along with the element that was left out of the first OCS, to the second OCS. The second OCS chooses one of these two elements; this choice becomes the final output of this timestep.

In Section~\ref{sect:single}, we analyze the performance of our three-way OCS for a special case first: we bound the probability that a certain element, say $u$, is never chosen for $k$ consecutive timesteps whose triples contain $u$. It is not that we require these $k$ timesteps to be consecutive in the original input: they need to be consecutive  in the subsequence of timesteps on which $u$ appeared in the triple.
By the definition of  two-way $\gamma$-OCS, the probability that the second two-way OCS never chooses $u$ is no greater than $(1/2)^j (1-\gamma)^{\max(j-1,0)}$ for some constant $\gamma$, where $j$ is the number of times $u$ was passed to the second OCS during those $k$ timesteps. Since this bound depends only on $j$, the question really reduces to determining (the probability distribution of) $j$.

In order for $u$ to be passed to the second OCS, it needs to be either left out of the first OCS or output by it.
It is easy to count how many times $u$ is left out of the first OCS: this follows a binomial distribution.
Therefore, the challenge is in counting the number of times $u$ is output by the first OCS. Unfortunately, its probability distribution highly depends on the actual \emph{input} to the first OCS, rather than the number of times $u$ is shown to the first OCS. Nonetheless, the following observation is crucial in coping with this difficulty: the probability distribution of the number of times Fahrbach et al.'s two-way OCS chooses $u$ is a unimodal symmetric distribution. Recall that the probability that the second OCS never chooses $u$ is bounded by $(1/2)^j (1-\gamma)^{\max(j-1,0)}$, which is ``nearly'' convex. Therefore, even though we cannot exactly calculate the probability distribution of $j$ without the full knowledge of the input, the above observation implies that we can instead use a ``flatter'' unimodal symmetric distribution in lieu of the actual distribution of $j$. Thanks to the near-convexity, this would give a valid upper bound on the probability. We formalize what a ``flatter'' distribution is by defining the notion of \emph{central dominance} as follows.
\begin{restatable}[Central Dominance]{definition}{centraldominance}
Given two discrete symmetric probability distributions $D_1$ and $D_2$ on $\{0, 1, \cdots, x\}$ whose probability mass functions are $p_1$ and $p_2$, respectively, we say $D_1$ \emph{centrally dominates} $D_2$ if there exists $z \in [0, \frac{x}{2}]$ such that, for any integer $y \in [ \frac{x}{2} - z, \frac{x}{2} + z]$, $p_1(y) \geq p_2(y)$, and for any integer $y \in [0, \frac{x}{2} - z) \cup (\frac{x}{2} + z, x]$, $p_1(y) \leq p_2(y)$.
\end{restatable}

We then construct our ``surrogate'' distribution that is centrally dominated by any possible probability distribution of $j$. This distribution depends only on the number of times the first OCS is given $u$. It is therefore much more amenable and allows us to obtain a bound on the probability that our three-way OCS never chooses $u$ from the given consecutive triples.

In Section~\ref{sect:general}, we generalize this bound to the non-consecutive case, i.e., a disjoint set of consecutive subsequences of timesteps containing $u$.
To obtain this bound, we perform a set of surgical operations that modify the input to the first OCS, which are designed to reduce negative correlation. These operations are inspired by those of Fahrbach et al.~\cite{fahrbach} that they used to obtain a similar generalization. In our case, however, we face a new obstacle: previously, it sufficed to bound only the probability that the two-way OCS never chooses a given element, since the output of that OCS was the final output. Our final output on the other hand is determined by the second OCS, and if our modification changes the output distribution of the first OCS, this may affect the output of the second OCS in an obscure way. We prove that a set of careful surgical operation can remove all correlations while ensuring that the bound is not affected. Once the correlations are removed, the general-case bound can be simply given as the product of our bounds from Section~\ref{sect:single} for single subsequences.

\begin{theorem}[simplified] \label{thm:mainsimpler}
Consider a set of $m$ disjoint consecutive subsequences of timesteps whose triples contain some element $u$ of the ground set. Let $k_1, \ldots, k_m$ be the lengths of these subsequences. The probability that our three-way OCS never chooses $u$ from these $m$ subsequences is at most\[
\prod_{i = 1}^m \left[ \left( \frac{2}{3} \right)^{k_i} (1 - \delta_1)^{\max(k_i - 1, 0)} (1 - \delta_2)^{\max(k_i - 2, 0)}\right]
,\] where $\delta_1 = 0.0309587$ and $\delta_2 = 0.0165525$.
\end{theorem}

Finally, in Appendices~\ref{sect:unweight} and~\ref{sect:weight}, we prove that our three-way OCS can be applied to edge-weighted and unweighted online bipartite matching problem.

\subsection{Related Work}
Introduced by Karp, Vazirani, and Vazirani~\cite{karp1990optimal}, the unweighted online bipartite matching has been intensively and extensively studied with alternative proofs~\cite{goel08online, birnbaum08on, devanur13randomized, eden2021economics} and under various settings including stochastic models~\cite{feldman2009online, manshadi2012online, haeupler2011online, mehta2015online, gamlath2019beating, huang21onlinestochastic}, fully online models~\cite{huang2020fully, huang2019tight, huang2020fully2}, and general arrival models~\cite{gamlath2019online}. The study of edge-weighted online bipartite matching problem was initiated by Kalyanasundaram \& Pruhs~\cite{kalyanasundaram1993online} and Khuller, Mitchell, \& Vazirani~\cite{khuller1994line}, who independently considered this problem under the metric assumption. Feldman, Korula, Mirrokni, Muthukrishnan, and P\'al~\cite{feldman2009freedisposal} first investigated the edge-weighted version on arbitrary weights with free disposal, and more thorough understanding of this problem was achieved by subsequent work~\cite{korula2013bicriteria, fahrbach}. Other variants and applications of the online bipartite matching have been studied as well, including  AdWords~\cite{mehta2007adwords, huang2020adwords},  vertex-weighted version~\cite{aggarwal2011online, huang2019online, gamlath2019beating}, stochastic or random arrival models~\cite{haeupler2011online, kesselheim2013optimal, brubach2016new, huang21onlinestochastic}, and the windowed version~\cite{ashlagi2019edge}. We refer interested readers to the survey of Mehta~\cite{mehta2013online}.

\subparagraph*{Recent related works.}
Recently, after the authors independently obtained the present results but prior to announcing them, the authors learned that two closely related papers were announced on arXiv~\cite{gao2021improved,blanc2021multiway}.

The other open question raised by Fahrbach et al.~\cite{fahrbach} than the one answered by this paper was to improve two-way OCS.
Gao, He, Huang, Nie, Yuan, and Zhong~\cite{gao2021improved} addresses this question by giving a novel automata-based OCS, successfully departing from the previous matching-based approach: their two-way OCS is a 0.167-OCS.
In addition to this, they also give an improved primal-dual analysis and a variant of two-way OCS specifically adapted for edge-weighted online bipartite matching, yielding a 0.519-competitve algorithm.
Finally, they consider a weaker relaxed notion of OCS called \emph{semi-OCS}, where the probability bound holds only for consecutive prefixes. They provide a multi-way version of this semi-OCS that leads to a 0.593-competitive algorithm for unweighted/vertex-weighted online bipartite matching.

Blanc and Charikar~\cite{blanc2021multiway} generalize Fahrbach et al.'s definition of OCS in two ways and give $m$-way OCS for any $m$. One of the two generalizations, called \emph{$(F,m)$-OCS}, gives the probability bound specified as a discrete function rather than a fixed-form formula such as $(1/2)^j(1-\gamma)^{\max(j-1,0)}$. The other is called \emph{continuous OCS}, which allows an element of the ground set to appear in a subset ``to a fraction'', where the probability bound is now specified as a continuous function. Their continuous OCS along with their improved primal-dual analysis gives a 0.5368-competitive algorithm for edge-weighted online bipartite matching. 

Considering these new results, an interesting question is whether the techniques from these papers and our independent result can together bring improvements in OCS or related problems such as online matching. In fact, since our framework treats the second two-way OCS as a black-box, any improved two-way OCS can be directly plugged into our three-way OCS. Combining the new two-way 0.167-OCS of Gao et al.~\cite{gao2021improved} with our results, for example, immediately yields a 0.5132-competitive algorithm for edge-weighted online bipartite matching.
\section{Preliminaries} \label{sect:prelim}
In this section, we present some notation, definitions, and previous results to be used throughout this paper.
Let us first introduce the definition of $n$-way OCS and two-way $\gamma$-OCS.
\begin{definition}[$n$-way OCS]
Given a ground set, an $n$-way OCS is an online algorithm that, at each iteration, takes a subset of size $n$ as the input and irrevocably chooses an element from the subset.
\end{definition}

For any element $u$ of the ground set, we say a subsequence of subsets containing $u$ is \emph{consecutive} if every subset containing $u$ that arrives between the first and last subsets of the subsequence is also in the subsequence. We also say that two or more subsequences of subsets containing $u$ are \emph{disjoint} if no two of these subsequences share a subset.

\begin{definition}[Two-way $\gamma$-OCS]\label{defn:gammaocs}
A two-way $\gamma$-OCS is a two-way OCS such that, for any element $u$ and a set of $m$ disjoint consecutive subsequences of pairs containing $u$ of lengths $k_1, \cdots, k_m$, the probability that $u$ never gets chosen by the OCS from any of the given subsequences is at most
$\prod_{i = 1}^m \left( \frac{1}{2} \right)^{k_i} (1 - \gamma)^{\max(k_i - 1, 0)}
$.
\end{definition}

Fahrbach et al.~\cite{fahrbach} presents two versions of two-way OCS with varying performance guarantees. We will use both in later sections. For the sake of completeness, we present a full description of the two-way $1/16$-OCS of Fahrbach et al. below.

\subparagraph*{$1/16$-OCS.}
We will define what is called the \emph{ex-ante graph} first. Although the algorithm does not need to explicitly construct this graph, it helps simplify the presentation.
The vertices of the ex-ante graph correspond to the input pairs. For each pair, say $\{u,v\}$, we introduce an edge between this pair and the immediately following pair that contains $u$, and likewise for $v$. For example, if an element $u$ appears at timesteps 2, 4, and 7, Pairs~2 and 4 are adjacent, and so are 4 and 7. This implies that every vertex has degree of at most 4. We annotate each edge with the common element that caused this edge. For example, if two pairs $i=\{u,v\}$ and $j=\{v,w\}$ are adjacent due to $v$, let $(i,j)_v$ denote the edge (and its annotation).

The algorithm samples exactly three random bits for each pair (or vertex). We will use them to define what is called the \emph{ex-post graph} and further determine the output of the algorithm. The ex-post graph is on the same set of vertices. The first random bit is used to determine if the vertex becomes a ``sender'' or a ``receiver''. If it becomes a sender, we use the second random bit to choose one of the two elements of the pair, and the sender will ``want'' to select the edge to the immediately following pair that shares the chosen element (if it exists). Similarly, if a vertex becomes a receiver, we use the second random bit to choose one element, and the receiver will ``want'' to select the edge to the immediately preceding pair that shares the element. Each edge of the ex-ante graph enters the ex-post graph if and only if both of its endpoints want to select the edge. Observe that the ex-post graph is a matching in the ex-ante graph.

Finally, the output of the algorithm is determined as follows. For every unmatched vertex, its output is determined solely by its third random bit. For each edge in the ex-post graph, we negatively correlate the choice of the two endpoints: we use the third random bit of the sender to determine the output of the sender, and the output of the receiver is determined so that the decision made for the shared element is the opposite. For example, if we have an edge annotated with $u$ in the ex-post graph and the sender did not choose $u$, we choose $u$ for the receiver; otherwise, we do not choose $u$ for the receiver. The third random bit of the receiver is discarded.

\begin{lemma}[Fahrbach et al.~\cite{fahrbach}]
This algorithm is a two-way $\frac{1}{16}$-OCS.
\end{lemma}

This algorithm also has the following useful property.
\begin{lemma} [Fahrbach et al.~\cite{fahrbach}] \label{lem:matchingprobbound}
Given a consecutive subsequence of pairs containing an element of length $k$ input to this algorithm, the probability that there does not exist any edges in the subgraph of the ex-post graph induced by the consecutive pairs is at most $(1 - 1/16)^{\max(k - 1, 0)}$.
\end{lemma}

The other version is a $\frac{13 \sqrt{13} - 35}{108}$-OCS.

\begin{lemma}[Fahrbach et al.~\cite{fahrbach}] \label{lem:gammaB}
There exists a two-way $\frac{13 \sqrt{13} - 35}{108}$-OCS.
\end{lemma}

In the \emph{unweighted online bipartite matching problem}, we are given a bipartite graph $G=(L \cup R, E)$ where we only know $L$ in advance. Each vertex $v \in R$ arrives one by one, and the edges adjacent with $v$ are only then revealed. Upon each arrival of $v$, we irrevocably decide whether we match $v$, and if so, to which exposed adjacent vertex in $L$ we match. The objective is to find a matching of the maximum size in $G$.

In the \emph{edge-weighted online bipartite matching problem with free disposal}, we are given a bipartite graph $G = (L \cup R, E)$ as well as an edge weight $w_{uv} \geq 0$ for each $(u, v) \in E$, where we only know $L$ in advance, again. Each vertex $v \in R$ arrives one at a timestep, and the edges adjacent with $v$ and their edge weights are disclosed at that time. Upon each arrival of $v$, we decide whether we match $v$, and if so, to which adjacent vertex in $L$ we match.
We remark that $v$ can be matched to a vertex $u$ that is already matched, after disposing of the edge incident with $u$ in the current matching (\emph{free disposal}).
The objective is to find a maximum-weight matching in $G$.

\section{Three-Way Online Correlated Selection} \label{sect:single}
In this section, we present our three-way OCS and analyze it by considering the special case where we are interested in a single consecutive subsequence of triples containing a common element. We will extend this to the general case in Section~\ref{sect:general}.

\subsection{Algorithm}
Let $\A$ and $\B$ be two-way OCS algorithms whose random choices are independent. In particular, we choose Fahrbach et al.'s $1/16$-OCS as $\A$ and $\frac{13 \sqrt{13} - 35}{108}$-OCS as $\B$.
Let $\gammaA:=1/16$ and $\gammaB:=\frac{13 \sqrt{13} - 35}{108}$.

Upon arrival of  a triple, say $\{u, v, w\}$, we choose a pair uniformly at random out of $\{ \{u, v\}, \{u, w\}, \{v, w\} \}$. We refer to this step as the \emph{random pair choice phase}. Without loss of generality, suppose that $\{u, v\}$ is chosen. We then input $\{u, v\}$ to $\A$ and let it choose one element from the pair. Let us say (without loss of generality again) that $u$ is returned by $\A$. Now we let $\B$ choose one element from $\{u, w\}$. The element chosen by $\B$ is the final output of this iteration.

\subsection{Overview of the Analysis}\label{ss:etaeta}
The goal of Section~\ref{sect:single} is to prove the following theorem.
\begin{theorem}\label{thm:etasimpler}
Consider a consecutive subsequence of timesteps whose triples contain some element $u$ of the ground set. Let $k$ be the length of the subsequence. The probability that our three-way OCS never chooses $u$ during these $k$ timesteps is at most
\[
\eta(k) := c_1 t_1^k + c_2 t_2^k - c_3 t_3^k - c_4 t_4^k
\]
for some $c_1 \approx 0.957795, \; c_2 \approx 0.176756, \; c_3 \approx 0.011047, \; c_4 \approx 0.131738, t_1 \approx  0.630024, \; t_2 \approx 0.599919,\; t_3 \approx 0.148345, \text{ and } t_4 = 0.3125$, which is bounded from above by
\[
\left( \frac{2}{3} \right)^k (1 - \delta_1)^{\max(k - 1, 0)} (1 - \delta_2)^{\max(k - 2, 0)},
\]
where $\delta_1 = 0.0309587$ and $\delta_2 = 0.0165525$.
\end{theorem}
In what follows, we will fix an arbitrary set of random choices made during the random pair choice phase, and condition on them.
Let $x$ be the number of pairs containing $u$ from the subsequence that are passed to $\A$. Observe that these pairs also form a consecutive subsequence to $\A$,
and that the number of $u$'s that are left out of the first OCS is $k-x$.

For $y = 0, \cdots, x$, let $p(x, y)$ be the (conditional) probability that $\A$ returns $y$ number of $u$'s from these pairs. Then, we can see that the (conditional) probability that $u$ never gets chosen from the given consecutive triples is bounded from above by
\begin{equation} \label{eq:condprobbound}
\sum_{y = 0}^x p(x, y) \left( \frac{1}{2} \right)^{k - x + y} (1 - \gammaB)^{\max(k - x + y - 1, 0)}
\end{equation}
by Lemma~\ref{lem:gammaB} and the fact that the pairs containing $u$ passed to $\B$ are also consecutive. To simplify the presentation, we introduce two shorthands: for a probability mass function $p'$ defined on $\{0, \cdots, x\}$, let
\begin{align}
\theta(x, p') & := \sum_{y = 0}^x p'(y) \cdot \left( \frac{1}{2} \right)^y \left( 1 - \gammaB \right)^{y - 1}; \text{ and} \label{eq:theta} \\
\theta'(x, p') & := \sum_{y = 0}^x p'(y) \cdot \left( \frac{1}{2} \right)^y \left( 1 - \gammaB \right)^{\max(y - 1, 0)}. \label{eq:thetaprime}
\end{align}
Observe that \eqref{eq:condprobbound} is equal to $\left( \frac{1 - \gammaB}{2} \right)^{k - x} \theta(x, p(x,\cdot))$ if $k < x$ and $\theta'(k, p(k,\cdot))$ otherwise.

We would like to bound \eqref{eq:condprobbound}, but unfortunately, $p(x, y)$ depends on the actual input to $\A$, not just $x$. Yet, we can circumvent this problem by exploiting the behavior of $\A$. Recall that $\A$ constructs the ex-post graph (which is a matching in the ex-ante graph) and negatively correlate the two endpoints of each edge in the ex-post graph. Therefore, when we consider the subgraph of the ex-post graph induced by the given consecutive pairs, we can observe that
\begin{itemize}
\item[-] for each edge in this subgraph, exactly one $u$ is chosen from its two endpoints, and
\item[-] if a vertex is isolated in this subgraph, $u$ is chosen with probability $1/2$, independently from the other vertices.
\end{itemize}
This observation implies that the probability distribution $p(x, \cdot)$ is a unimodal symmetric distribution. Moreover, the more likely $\A$ puts an edge in the ex-post graph, the pointier the distribution would be. Using this property, we will construct an imaginary probability distribution $\{p^*(x, y)\}_{y = 0, \cdots, x}$ such that
\begin{itemize}
\item[-]  $p^*(x,y)$ is flatter than (or, formally, centrally dominated by) any distribution $p$ that results from $\A$ (see the proof of Lemma~\ref{lem2}); and
\item[-] $p^*(x,y)$ only depends on $x$,  not the input itself.
\end{itemize}

We will further demonstrate that, in \eqref{eq:theta} and \eqref{eq:thetaprime}, substituting $p'$ with  a centrally dominated distribution would only overestimate \eqref{eq:theta} and \eqref{eq:thetaprime} (Lemma~\ref{lem1}). Intuitively speaking, this is because $(1/2)^y (1 - \gammaB)^{y - 1}$ and $(1/2)^y (1-\gammaB)^{\max(y - 1, 0)}$ are (nearly) convex functions.

Thus, the probability that our three-way OCS never chooses $u$ from the given consecutive triples can finally be bounded from above by
\begin{equation}
\eta(k) := \sum_{x = 0}^k \binomprob \left( k, x, \frac{2}{3} \right) \left[ \sum_{y = 0}^x p^* (x, y) \left(\frac{1}{2} \right)^{k - x + y} (1 - \gammaB)^{\max(k - x + y - 1, 0)} \right], \label{eq:eta}
\end{equation}
where $\binomprob(k, x, r)$ represents the probability of the binomial distribution for $x$ successes out of $k$ trials with probability $r$ (Lemma~\ref{lem:singleconsecbound}).

\subsection{Using a Centrally Dominated Distribution}
Recall that $k$ denotes the length of the subsequence, i.e., the number of triples, and $x$ denotes the number of pairs containing $u$ passed to $\A$. Let $y$ be the number of $u$'s returned by $\A$; then, $\B$ receives a consecutive subsequence of pairs containing $u$ of length $k - x + y$.

We remark that the probability that $u$ is never chosen by $\B$, conditioned on the event that $\A$ takes $x$ pairs containing $u$ and outputs $y$ number of $u$'s, can be bounded by
$
\left( \frac{1}{2} \right)^{k - x + y}\cdot  (1 - \gammaB)^{\max(k - x + y - 1, 0)}
$
from Lemma~\ref{lem:gammaB}. Note that this bound depends only on the lengths $k$, $x$, and $y$, not the actual input to $\B$.

For $\A$, on the other hand, we need to calculate the probability that exactly $y$ number of $u$'s are chosen by $\A$ from the given subsequence of pairs. However, this probability distribution highly depends on the input to $\A$; therefore, we cannot conveniently fix a distribution parameterized by length $x$ anymore. Nonetheless, we will show that there exists a bounding distribution, depending only on $x$, such that it yields a valid upper bound on the probability that $u$ is never chosen by our entire algorithm from the given subsequence of triples.

\subsubsection{Central Dominance}
To construct the bounding distribution, we need the notion of central dominance. Recall the definition.
\centraldominance*

Intuitively speaking, this definition indicates how flat a symmetric probability distribution is. Following is the lemma that formalizes why a flat distribution helps bound the probability that $u$ never gets chosen from our three-way OCS. The definitions of $\theta(\cdot, \cdot)$ and $\theta'(\cdot, \cdot)$ can be found in \eqref{eq:theta} and \eqref{eq:thetaprime}, respectively.
We defer its proof to Appendix~\ref{app:dproofs}. It is easy to intuitively see that the first half of the lemma should hold; yet, $\theta'$ is only \emph{nearly} convex and requires some work.
\begin{restatable}{lemma}{convexbound} \label{lem1}
Suppose we are given two discrete symmetric distributions $D_1$ and $D_2$ on $\{0, 1, \cdots, x\}$ whose probability mass functions are $p_1$ and $p_2$, respectively. If $D_1$ centrally dominates $D_2$, we have $\theta(x, p_1) \leq \theta(x, p_2)$ and $\theta'(x, p_1) \leq \theta'(x, p_2)$.
\end{restatable}

\subsubsection{Bounding Distribution}
In this section, we construct the bounding distribution. As was noted earlier, $\A$ negatively correlates the endpoints of every edge of the ex-post graph.
Therefore, we can write the probability $p(x, y)$ that the two-way OCS chooses precisely $y$ number of $u$'s out of the given $x$ consecutive pairs if we are given $q \in \R_+^{\floor{x/2} + 1}$ defined as follows: for $i = 0, \cdots, \floor{\frac{x}{2}}$, $q_i$ is the probability that exactly $i$ edges of the ex-post graph have both endpoints in the given subsequence. Note that $ \sum_{i = 0}^\floor{x/2} q_i = 1$. We have
\[
p(x, y) = q_0 \cdot \binom{x}{y} \left( \frac{1}{2} \right)^{x} + q_1 \cdot \binom{x - 2}{y - 1} \left( \frac{1}{2} \right)^{x - 2} + \cdots +  q_\floor{x/2} \cdot  \binom{x - 2 \floor{x/2}}{y - \floor{x/2}} \left( \frac{1}{2} \right)^{x - 2{\floor{x/2}}},
\]
where ${b \choose c} := 0$ if $b < c$ or $c < 0$.

Note that
$q$ (and therefore in turn $p$) depends on the actual input to $\A$. Given any probability vector $q \in \R_+^{\floor{x/2} + 1}$, let $D(q)$ (or $D(q_0, \cdots, q_{\floor{x/2}})$) denote the probability distribution $\{p(x, y)\}_{y=0,\ldots,x}$. It is easy to verify that, for all $q$, $D(q)$ is a valid probability distribution that is symmetric on $y = \frac{x}{2}$.

Now we construct the bounding distribution. For $x = 0, \cdots, k$, let $\alpha_x := (1 - \gammaA)^{\max(x - 1, 0)}$ for $\gammaA = 1/16$. We choose $D(\alpha_x, 1 - \alpha_x, 0, \cdots, 0)$ as our bounding distribution. Let $p^*(x, \cdot)$ be the probability mass function of the bounding distribution, i.e., for each $y = 0, \cdots, x$,
\[
p^*(x, y) := \left\{ \begin{array}{ll} \alpha_x \left( \frac{1}{2} \right)^x , & \text{if } y = 0 \text{ or } x, \\ \alpha_x {x \choose y} \left( \frac{1}{2} \right)^x + (1 - \alpha_x) {x-2 \choose y-1} \left( \frac{1}{2} \right)^{x - 2}, & \text{otherwise.}  \end{array} \right.
\]

Following is a key lemma to show that $D(\alpha_x, 1 - \alpha_x, 0, \cdots, 0)$ is a good choice of the centrally dominated distribution.
\begin{lemma} \label{lem2}
For any $x = 0, \cdots, k$, and any valid probability distribution $D(q)$ followed by $p(x, \cdot)$ for some probability vector $q \in \R_+^{\floor{x/2} + 1}$ that the two-way OCS can have when given a consecutive subsequence of length $x$, we have $ \theta(x, p(x, \cdot)) \leq \theta(x, p^*(x, \cdot)) $ and $ \theta'(x, p(x, \cdot)) \leq \theta'(x, p^*(x, \cdot)) $.
\end{lemma}
\begin{proof}
For $x = 0, 1$, it is easy to see that $p(x, y) = p^*(x, y)$ for every possible $y$ since $q_0 = 1 = \alpha_x$. For $x \geq 2$, we construct intermediate distributions as follows:
\begin{align*}
D_0 &:= D(q_0, 1-q_0, 0, \cdots, 0),\\
D_1 &:= D(q_0, q_1, 1- q_0 - q_1, \cdots, 0),\\
& \cdots, \\
D_{\floor{x/2}-1} &:= D(q_0, q_1, q_2, \cdots, \textstyle 1- \sum_{i = 0}^{\floor{x/2} - 1}q_i) = D(q).
\end{align*}
We claim that $D_0$ centrally dominates $D(\alpha_x, 1-\alpha_x, 0, \cdots, 0)$ and, for each $j = 1, \cdots ,\floor{\frac{x}{2}}-1$, $D_j$ centrally dominates $D_{j - 1}$. Then, by repeatedly applying Lemma~\ref{lem1}, we can prove the lemma.

Let us first prove that $D_0$ centrally dominates $D(\alpha_x, 1-\alpha_x, 0, \cdots, 0)$. Since both distributions are symmetric, it is sufficient for us to consider $y \in [0, \frac{x}{2}]$. By Lemma~\ref{lem:matchingprobbound}, we have $q_0 \leq \alpha_x$. If $q_0 = \alpha_x$, $D_0$ is equivalent to $D(\alpha_x, 1-\alpha_x, 0, \cdots, 0)$, completing the proof. We now assume $q_0 < \alpha_x$. If $y = 0$, $D_0$ has the smaller probability. This implies that there must exist a point in $[1, \frac{x}{2}]$ where $D_0$ has the greater probability since both distributions are valid.

Let $y^*$ be the smallest such point. We show that for any integer $y \in [y^*, \frac{x}{2}]$, $D_0$ has the greater probability. Let
$
g(y) := (q_0 - \alpha_x) {x \choose y} \left( \frac{1}{2} \right)^x + (\alpha_x - q_0) {x-2 \choose y-1} \left( \frac{1}{2} \right)^{x-2}
$
be the subtraction of the probability of $D_0$ from that of $D(\alpha_x, 1 - \alpha_x, 0, \cdots, 0)$ at point $y$. It suffices to prove that $g(y)$ is non-decreasing on $[y^*, \frac{x}{2}] \subseteq [1, \frac{x}{2}]$. By rewriting the formula, we have
$
g(y) = (\alpha_x - q_0){x \choose y} \left( \frac{1}{2} \right)^x \left( \frac{y(x - y)}{4x(x-1)} - 1 \right)
$,
and it is not hard to see that the function is increasing on $[1, \frac{x}{2}]$.

A similar argument can be applied to show that, for each $j = 1, \cdots, \floor{\frac{x}{2}} - 1$, $D_j$ centrally dominates $D_{j - 1}$. Indeed, if $q_j = 1 - \sum_{i = 0}^{j - 1}q_i$, $D_j$ is equivalent to $D_{j - 1}$. We thus assume that $q_j < 1 - \sum_{i = 0}^{j - 1}q_i$. Observe that $D_j$ has the same probabilities as $D_{j - 1}$ for $y = 0, \cdots, j - 1$, and has the smaller probability for $y = j$; hence, there exists a point in $\left[ j + 1, \frac{x}{2} \right]$ where $D_j$ has the greater probability, and let $y^*$ be the smallest such value. Again, let $g(y)$ be the difference obtained by subtracting the probability of $D_{j - 1}$ from that of $D_j$ at point $y$. We can write
\begin{align*}
g(y) &= (1 - \sum_{i = 0}^j q_i) \left[ {x - 2j - 2 \choose y - j - 1} \left( \frac{1}{2} \right)^{x - 2j - 2} - {x - 2j \choose y - j} \left( \frac{1}{2} \right)^{x - 2j}\right] \\
	& = (1 - \sum_{i = 0}^j q_i) {x - 2j \choose y - j} \left( \frac{1}{2} \right)^{x - 2j} \left[ \frac{(y - j) (x - y - j)}{4(x - 2j) (x - 2j - 1)} - 1 \right],
\end{align*}
implying that $g(y)$ is increasing on $\left[ y^*, \frac{x}{2}\right] \subseteq \left[j + 1, \frac{x}{2}\right]$.
\end{proof}

We are now ready to bound the probability that our three-way OCS never chooses the element $u$ from the consecutive subsequence of triples of length $k$. Let $\binomprob(k, x, r)$ be the probability mass function of the binomial distribution for $x$ successes out of $k$ trials with probability $r$, i.e.,
$
\binomprob(k, x, r) := {k \choose x} r^x (1 - r)^{k - x}
$.
Finally, we define $\eta(k)$ as follows and it will be the desired bound.
\[
\eta(k) := \sum_{x = 0}^k \binomprob \left( k, x, \frac{2}{3} \right) \left[ \sum_{y = 0}^x p^* (x, y) \left(\frac{1}{2} \right)^{k - x + y} (1 - \gammaB)^{\max(k - x + y - 1, 0)} \right].
\]

\begin{lemma} \label{lem:singleconsecbound}
Given a consecutive subsequence of triples containing $u$ of length $k$, the probability that our three-way OCS never chooses $u$ from the subsequence is at most $\eta(k)$.
\end{lemma}
\begin{proof}
Conditioned on the pairs selected by the random pair choice phase, let $x$ be the number of pairs containing $u$ inserted into $\A$. Recall that these pairs form a consecutive subsequence. Let $p(x, y)$ denote the probability that $\A$ returns $y$ number of $u$'s from this subsequence. Note that $p$ follows $D(q)$ for some $q$.

Note that, if $\A$ returns $y$ number of $u$'s, $\B$ eventually takes $k - x + y$ consecutive pairs that originate from the given consecutive triples. By Lemma~\ref{lem:gammaB}, we can see that the probability that $u$ is never selected from the given consecutive subsequences of triples is bounded from above by
$$
\sum_{y = 0}^x p(x, y) \left( \frac{1}{2} \right)^{k - x + y} (1 - \gammaB)^{\max(k - x + y - 1, 0)},
$$
which can be rewritten as $
 \left( \frac{1 - \gammaB}{2} \right)^{k - x} \theta(x, p(x, \cdot))$ if $x < k$ or $ \theta'(k, p(x, \cdot)) $ if $ x = k $.
In any case, by Lemma~\ref{lem2}, it can be further bounded by
$$
\sum_{y = 0}^x p^*(x, y) \left( \frac{1}{2} \right)^{k - x + y} (1 - \gammaB)^{\max(k - x + y - 1, 0)}.
$$
It is noteworthy that this value depends only on the number $x$ of $u$'s selected by the random pair choice phase. Observe that the probability that $x$ pairs containing $u$ are chosen from $k$ consecutive triples by the random pair choice phase is exactly $\binomprob(k, x, 2/3)$, completing our proof.
\end{proof}

It still remains to calculate the bounds on $\eta(k)$ in the same form as stated in Theorem~\ref{thm:etasimpler}. See Appendix~\ref{app:caleta} for the detailed calculation.
\section{General Bound} \label{sect:general}
We now extend our discussion to the general case and provide a bound for a set of disjoint consecutive subsequences. The following theorem states this bound.
\begingroup
\def\thetheorem{\ref{thm:mainsimpler}}
\begin{theorem}[restated] \label{thm:mainrestate}
Consider a set of $m$ disjoint consecutive subsequences of triples containing an element $u$. Let $k_1, \ldots, k_m$ be the lengths of these subsequences. The probability that our three-way OCS never chooses $u$ from these $m$ subsequences is at most $\prod_{i = 1}^m \eta(k_i)$.
\end{theorem}
\addtocounter{theorem}{-1}
\endgroup

Let $\kvect = (k_1, \cdots, k_m)$  be a vector whose $i$-th entry is the length of the $i$-th subsequence. In what follows, we fix (and condition upon) the choices made by the \emph{random pair choice phase}. Let $\xvect := (x_1, \cdots, x_m)$ be the (now constant) vector whose $i$-th entry represents the number of  pairs containing $u$  from the $i$-th subsequence that are passed to $\A$. For $\yvect := (y_1, \cdots, y_m) \leq \xvect$, let $p(\xvect, \yvect)$ be the conditional probability that $\A$ chooses $y_i$ number of $u$'s from the $i$-th subsequence of pairs for all $i$. Finally, let $p_0$ be the probability that our three-way OCS never chooses $u$. By Lemma~\ref{lem:gammaB}, we can observe that $p_0$ is no greater than
\begin{equation} \label{eq:jointprobobj}
\sum_{\yvect \leq \xvect} p(\xvect, \yvect) \prod_{i = 1}^m \left( \frac{1}{2} \right)^{k_i - x_i + y_i} (1 - \gammaB)^{\max(k_i - x_i + y_i - 1, 0)}.
\end{equation}
To prove the desired bound, we will first present how we can modify the input to $\A$ without ever decreasing \eqref{eq:jointprobobj}. After finitely many modifications, we will be able to ``decouple'' the random choices across different subsequences.

Recall that each pair inserted into $\A$ takes three independent random bits. The first random bit determines if the pair is a sender or a receiver. The second random bit decides which adjacent vertex to correlate. The third random bit selects an element to be output for this pair, unless the pair becomes matched as a receiver. Note that, with the first two random bits, the ex-post graph can be determined.

For each vertex $j$ containing $u$ in the ex-ante graph, let $\pred{j, u}$ and $\pred{j, -u}$ be the immediate predecessor of $j$ linked by $u$ and by the other element than $u$, respectively. Without loss of generality, let $v$ be the other element of $j$ than $u$. If we have $j'$ and $j$, residing in different subsequences and $j'$ appearing before $j$, such that
\begin{enumerate}[(A)]
\item~\label{badtype1} $j' = \pred{j, u}$ or $j' = \pred{j, -u}$ (or both); or
\item~\label{badtype2} there exists $\jhat$ such that
\begin{itemize}
\item[-] $\jhat = \pred{j', u}$ or $\jhat = \pred{j', -u}$,
\item[-] $\jhat = \pred{j, -u}$, and
\item[-] $\jhat$ is not contained in any subsequence,
\end{itemize}
\end{enumerate}
we call them \emph{violations}. In particular, the first type of violations are called Type~\ref{badtype1} violations and the other Type~\ref{badtype2}. Our goal is to modify the input so that there are no violations while \eqref{eq:jointprobobj} never decreased.

For notational simplicity, we let
$
\phi(y) := \left( \frac{1}{2} \right)^{k - x + y} (1 - \gamma_\B)^{\max(k - x + y - 1, 0)}
$
for all $y = 0, \cdots, x$.
Observe that $\phi(y)$ is decreasing over $y$. We can now consider \eqref{eq:jointprobobj} as the expected value of $\prod_{i=1}^m \phi(y_i)$.

\subparagraph*{Removing Type~\ref{badtype1} Violations.}
Let us first consider the case when $j' = \pred{j, u} = \pred{j,  -u}$. We build the new input to $\A$ by inserting $\jbar_1 = (u, \star)$ and $\jbar_2 = (v, \diamond)$ right before $j$ is input where $\star$ and $\diamond$ are elements appearing nowhere else and $\star \not= \diamond$.

Let us fix some random bits of $\A$: one can think of this as further conditioning on those bits. In particular, we will fix the first two random bits of every pair except for $\jbar_1$ and $\jbar_2$. 

Suppose for now that the random bits dictate that neither $(j', j)_u$ nor $(j',j)_v$ is present in the ex-post graph of the original input. We claim that the probability distribution $p$ conditioned on this event stays the same even after the modification and therefore the ``contribution'' to the expectation \eqref{eq:jointprobobj} is not affected either (or in other words, the conditional expectation remains equal).

We will show a stronger statement that the marginal probability distribution of the pairs in the subsequences stays the same. (Then the probability distribution $p$ will automatically be the same.) Note that the ex-post graph of the two inputs will look almost identical, except for possible addition of some edges incident with $\jbar_1$ or $\jbar_2$. Therefore, the output choice of every pair in the subsequences will be determined by the same random bit in both inputs, except that, it may be the case that $j$ was not adjacent with any pair in the ex-post graph of the original input but becomes a receiver of $\jbar_1$ (or $\jbar_2$) after the modification. In this case, the output choice of $j$ was determined by the third random bit of $j$ before but by that of $\jbar_1$ (or $\jbar_2$) now. However, the only pair in the subsequence whose output choice is determined by the third random bit of $\jbar_1$ (or $\jbar_2$) is $j$. Moreover, the third random bits of the pairs are i.i.d.; hence, the marginal distribution of the pairs in the subsequences stays the same, as was claimed.

Now consider the remaining case where $j'$ and $j$ are adjacent in the ex-post graph of the original input. In this case, we will show that the conditional expectation cannot decrease (as opposed to staying the same). To ease the argument, we will fix more random bits (and argue that the conditional expectation cannot decrease in all cases): we fix the third random bits of all pairs, except for $j'$, $j$, $\jbar_1$, and $\jbar_2$. Let $z'$ be the number of $u$'s chosen from the subsequence containing $j'$, except for $j'$ itself, in the original input. (Since we did not fix the third random bit of $j'$, we cannot determine what the output choice of $j'$ is.) Similarly, let $z$ be the number of $u$'s chosen from the subsequence containing $j$, except for $j$ itself.

In the original input, note that the output choice of $j'$ and $j$ are both determined by the third random bit of $j'$. That is, with probability $1/2$, $z'$ and $z+1$ respectively are the number of $u$'s chosen from the subsequences containing $j'$ and $j$, and with probability $1/2$, $z'+1$ and $z$ are. In the modified input, however, while we do not know whether $j'$ is adjacent with $\jbar_1$ or $\jbar_2$ in the ex-post graph, $\jbar_1$ and $\jbar_2$ can each be adjacent with at most one pair in the ex-post graph. Therefore, in the marginal distribution of the pairs in the subsequences, the output choice of $j'$ is independent from all other pairs, including $j$. This shows that the number of $u$'s chosen from the subsequence containing $j'$ in the modified input is $z'$ with probability $1/2$ and $z'+1$ with probability $1/2$, and independently from that, the number of $u$'s from the subsequence containing $j$ is $z$ with probability $1/2$ and $z+1$ with probability $1/2$.

Let us now calculate the increase of the conditional expectation of $\prod_{i=1}^m\phi(y_i)$, but since we are only interested in its sign, we will ignore the common terms, i.e., $\phi(y_i)$'s for the subsequences other than ones containing $j$ and $j'$.
Note that
$
\frac{1}{4} [ \phi(z') \phi(z) + \phi(z') \phi(z + 1)  + \phi(z' + 1) \phi(z) + \phi(z' + 1) \phi(z + 1) ]
- \frac{1}{2}[ \phi(z') \phi(z + 1) + \phi(z' + 1) \phi(z) ]
= \frac{1}{4}(\phi(z') - \phi\left(z' + 1\right))(\phi(z) - \phi(z + 1))  \geq 0,
$
where the inequality follows from the fact that $\phi(y)$ is decreasing over $y$. This shows our claim.

Let us now consider the case where $j' = \pred{j, u} \not= \pred{j, -u}$ or vice versa. In this case, we insert $\jbar = (u, \star)$ or $\jbar = (v, \star)$ right before $j$, where $\star$ again is a unique element. A similar argument shows that we can show that \eqref{eq:jointprobobj} can only increase in the new input.

\subparagraph*{Omitted proof.}
We still need to remove Type~\ref{badtype2} violations, but we will defer these details to Appendix~\ref{sect:extend}. Intuitively speaking, we can further modify the input so that, eventually, no two pairs from different subsequences are in the same connected component of the ex-ante graph. Since $\A$ negatively correlates only those pair of vertices that are adjacent in the ex-post graph, this shows that the random choices within different subsequences can be ``completely decoupled''. This allows us to ``independently'' apply the bounds for Section~\ref{sect:single}, proving the theorem. See Appendix~\ref{sect:extend} for details.

\section{Application to Online Bipartite Matching} \label{sect:apply}
Our three-way OCS can be applied to online bipartite matching problems. 
While the algorithm and the factor-revealing LP had to be generalized to a higher dimension, our algorithm and analysis are still analogous to Fahrbach et al.~\cite{fahrbach}.
In the interest of space, we defer the detailed presentation to Appendices~\ref{sect:unweight} and~\ref{sect:weight}.
\begin{theorem}\label{thm:uwm}
There exists a $0.5096$-competitive algorithm for unweighted online bipartite matching.
\end{theorem}
\begin{theorem}\label{thm:wwmm}
There exists a $0.5093$-competitive algorithm for edge-weighted online bipartite matching with free disposal.
\end{theorem}

\bibliography{lit}
\appendix

\section{Proof of Lemma~\ref{lem1}} \label{app:dproofs}

\convexbound*
\begin{proof}
If $x = 0, 1$, it is trivial since there is only one possible symmetric distribution. Now we assume that $x \geq 2$. Given a discrete symmetric distribution with a probability mass function $p$, let us consider the following operation: For some $w, z \in [0, \frac{x}{2}]$ such that $w > z$, we decrease $p(\frac{x}{2} - z)$ and $p(\frac{x}{2} + z)$ by some $\epsilon > 0$, and increase $p(\frac{x}{2} - w)$ and $p(\frac{x}{2} + w)$ by $\epsilon$, instead. We can observe that $p$ still forms a valid probability distribution, and changes $\theta(x, p)$ by
\[
\Delta\theta(x,p) = \epsilon \left(\frac{1}{2}\right)^{x/2} (1-\gammaB)^{x/2 - 1} \left[ \left(\frac{1 - \gammaB}{2}\right)^{w} + \left( \frac{1 - \gammaB}{2}\right)^{-w} -  \left(\frac{1 - \gammaB}{2}\right)^{z} - \left(\frac{1 - \gammaB}{2}\right)^{-z} \right].
\]
Let $\alpha := (1-\gammaB) / 2$ and $f(t) := \alpha^t + \alpha^{-t}$. Since $f'(t) = \ln \alpha \cdot (\alpha^t - \alpha^{-t})$, we can verify that $f(t)$ is increasing on $t \geq 0$. Hence, we can see that
\[
\Delta \theta(x, p) = \epsilon \left( \frac{1}{2} \right)^{x/2} (1-\gammaB)^{x/2 - 1} (f(w) - f(z)) \geq 0.
\]
Since $D_1$ centrally dominates $D_2$, we can transform $p_1$ into $p_2$ by transferring the mass through a few times of the above operation with proper $w$'s, $z$'s, and $\epsilon$'s. For each application, $\theta(x, p)$ only increases, yielding that $\theta(x, p_1) \leq \theta(x, p_2)$.

Proving the other statement is more subtle. Suppose we transform $p_1$ into $p_2$ through the same method. If we increase the probabilities of $\frac{x}{2} - w$ and $\frac{x}{2} + w$ where $w < \frac{x}{2}$, we can apply to the above analysis since the change of $\theta'(x, p)$ is exactly the same as $\Delta \theta(x, p)$.

It remains to consider when the probabilities of $0$ and $x$ get increased by $\epsilon$. Assume that we instead decrease the probabilities of $\frac{x}{2} - z$ and $\frac{x}{2} + z$ where $0 \leq z \leq \frac{x}{2} - 1$. The change of $\theta'(x, p)$ would be
\[
\Delta \theta'(x,p) := \epsilon \left[ 1 + \left( \frac{1}{2} \right)^x (1-\gammaB)^{x-1} - \left( \frac{1}{2} \right)^{x/2 - z} (1-\gammaB)^{x/2 - z -1} - \left( \frac{1}{2} \right)^{x/2 + z} (1-\gammaB)^{x/2 + z -1} \right].
\]
Let $\beta := (1/2)^{x/2}(1-\gammaB)^{x/2 - 1}$ and $g(z) := (2/(1-\gammaB))^z$. We then rewrite $\Delta\theta'(x,p)$ as
\[
\Delta \theta'(x,p) = \epsilon \left[ 1 + \beta^2(1-\gammaB) - \beta ( g(z) + 1/g(z)) \right].
\]
The partial derivative of $\Delta \theta'(x,p)$ with respect to $z$ is 
\[
\frac{\partial \Delta \theta'(x, p) }{\partial z} = - \epsilon \beta \ln \left(\frac{2}{1-\gammaB}\right) \left[ \left(\frac{2}{1-\gammaB}\right)^z - \left(\frac{1-\gammaB}{2}\right)^z \right].
\]
Note that $\frac{\partial \Delta \theta'(x, p)}{\partial z} \leq 0$ for $z \geq 0$. Thus, we can bound $\Delta \theta'(x,p)$ from below by plugging $z = \frac{x}{2} - 1$, i.e.,
\[
\Delta \theta'(x,p) \geq \epsilon \left[ \frac{1}{2} - \frac{1 + \gammaB}{2} \left(\frac{1}{2}\right)^{x - 1} (1 - \gammaB)^{x - 2} \right].
\]
Since $x \geq 2$, we have
\[
\Delta \theta'(x,p) \geq \epsilon \left[ \frac{1}{2} - \frac{1 + \gammaB}{4} \right] \geq 0,
\]
where the last inequality comes from the fact that $\gammaB < 1$.
\end{proof}

\section{Calculating \texorpdfstring{$\eta(k)$}{eta(k)}} \label{app:caleta}
In this appendix, we calculate the desired bounds on $\eta(k)$.

\begin{lemma}
We have
\[
\theta(x, p^*(x, \cdot)) = \left\{ \begin{array}{ll}
\frac{8}{(3 - \gammaB)^2} + \frac{(1 + \gammaB)^2}{(1-\gammaA)(1 - \gammaB)(3-\gammaB)^2} - \frac{ \gammaA (1 + \gammaB)^2}{(1-\gammaA) (1-\gammaB) (3-\gammaB)^2}, & \text{ if } x = 0, \\
\frac{8}{(3-\gammaB)^2} \left( \frac{3 - \gammaB}{4} \right)^x +  \frac{(1 + \gammaB)^2}{(1 - \gammaA) (1 - \gammaB) (3-\gammaB)^2} \left( \frac{(3 - \gammaB)(1 - \gammaA)}{4} \right)^x, & \text{ otherwise.}  \end{array} \right.
\]
\end{lemma}
\begin{proof}
For $x > 0$, we have $ \alpha_x = (1 - \gammaA)^{x - 1}$. We break $\theta(x, p^*(x, \cdot))$ into two summations:
\begin{align*}
\theta(x, p^*(x, \cdot))  = & \alpha_x \left[ \sum_{y = 0}^x {x \choose y} \left( \frac{1}{2} \right)^x \left( \frac{1}{2} \right)^y (1 - \gammaB)^{y-1}  \right] \\
& + ( 1 - \alpha_x )  \left[ \sum_{y = 1}^{x-1} {x - 2 \choose y - 1} \left( \frac{1}{2} \right)^{x-2} \left( \frac{1}{2} \right)^y (1 - \gammaB)^{y-1}  \right].
\end{align*}
For the first summation, we can rewrite it as follows:
\[
\frac{\alpha_x}{2^x (1-\gammaB)} \sum_{y = 0}^x {x \choose y} \left(\frac{1 - \gammaB}{2}\right)^y = \frac{\alpha_x}{1 - \gammaB} \left(\frac{3-\gammaB}{4}\right)^x,
\]
where the equation follows from the binomial theorem. Similarly, the second summation can also be rephrased as follows with $y' := y - 1$:
\[
\frac{1 - \alpha_x}{2^{x - 1}} \sum_{y' = 0}^{x - 2} {x-2 \choose y'} \left(\frac{1 - \gammaB}{2}\right)^{y'} = \frac{1 - \alpha_x}{2} \left(\frac{3 - \gammaB}{4}\right)^{x - 2}.
\]
Summing up these values, we obtain
\begin{align*}
\theta(x, p^*(x, \cdot)) & = \left( \frac{3 - \gammaB}{4} \right)^x \left[ \frac{\alpha_x}{1 - \gammaB} + \frac{1 - \alpha_x}{2} \left( \frac{4}{3-\gammaB} \right)^2 \right] \\
&=  \left( \frac{3 - \gammaB}{4} \right)^x \left[ \frac{8}{(3 - \gammaB)^2} + \frac{(1 + \gammaB)^2}{(1 - \gammaA) (1 - \gammaB) (3-\gammaB)^2} (1-\gammaA)^x \right] \\
& = \frac{8}{(3-\gammaB)^2} \left( \frac{3 - \gammaB}{4} \right)^x +  \frac{(1 + \gammaB)^2}{(1 - \gammaA) (1 - \gammaB) (3-\gammaB)^2} \left( \frac{(3 - \gammaB)(1 - \gammaA)}{4} \right)^x
\end{align*}

If $x = 0$, we have $\theta(x, p^*(x, \cdot)) = 1 / (1 - \gammaB)$. Observe that
\[
\frac{8}{(3 - \gammaB)^2} + \frac{(1 + \gammaB)^2}{(1-\gammaA)(1 - \gammaB)(3-\gammaB)^2} - \frac{ \gammaA (1 + \gammaB)^2}{(1-\gammaA) (1-\gammaB) (3-\gammaB)^2} = \frac{1}{1 - \gammaB}.
\]
\end{proof}
 
\begin{lemma}
We have
\[
\theta'(x, p^*(x, \cdot)) = \left\{ \begin{array}{ll} 1, & \text{ if } x = 0, \\ \theta(x, p^*(x, \cdot)) - \frac{\gammaB}{(1 - \gammaA)(1-\gammaB)} \left(\frac{1 - \gammaA}{2} \right)^x, & \text{ otherwise.}  \end{array} \right.
\]
\end{lemma}
\begin{proof}
If $x = 0$, it is easy to see that $\theta'(x, p^*(x, \cdot)) = 1$. Otherwise, observe that
\[
\theta'(x, p^*(x, \cdot)) = \theta(x, p^*(x, \cdot)) - \frac{\gammaB}{1 - \gammaB} p^*(x, 0) = \theta(x, p^*(x, \cdot)) - \frac{\gammaB}{(1 - \gammaA)(1 - \gammaB)} \left( \frac{1 - \gammaA}{2} \right)^x.
\]
\end{proof}

Let
\[
c_1 := \frac{8}{(3-\gammaB)^2},\; c_2 := \frac{(1 + \gammaB)^2}{(1-\gammaA) (1 - \gammaB) (3-\gammaB)^2},
\]
\[
c_3 := \frac{\gammaA (1+\gammaB)^2}{(1-\gammaA) (1 - \gammaB) (3-\gammaB)^2}, \; c_4 := \frac{\gammaB}{(1-\gammaA)(1 - \gammaB)},
\]
\[
t_1 := \frac{2 - \gammaB}{3},\; t_2 := \frac{4 - 3\gammaA - 2\gammaB + \gammaA \gammaB}{6},\;
t_3 := \frac{1-\gammaB}{6}, \text{ and } t_4 := \frac{1-\gammaA}{3}.
\]
For $\gammaA = 1/16, \gammaB = \frac{13 \sqrt{13} - 35}{108}$, we have
\[
c_1 \approx 0.957795, \; c_2 \approx 0.176756, \; c_3 \approx 0.011047, \; c_4 \approx 0.131738,
\]
\[
t_1 \approx  0.630024, \; t_2 \approx 0.599919,\; t_3 \approx 0.148345, \text{ and } t_4 = 0.3125.
\]    
\begin{lemma} \label{lem:unweighted-eta}
We have
\[
\eta(k) = \left\{ \begin{array}{ll}
1, & \text{ if } k = 0, \\
c_1 t_1^k + c_2 t_2^k - c_3 t_3^k - c_4 t_4^k, & \text { otherwise.}
\end{array} \right.
\]
Moreover, for $\gammaA = 1/16, \gammaB =  \frac{13 \sqrt{13} - 35}{108}$, $\eta(k)$ is decreasing.
\end{lemma}
\begin{proof}
If $k = 0$, it is easy to see that $\eta(k) = 1$. For $k > 0$, we have
\begin{align*}
\eta(k) = & \sum_{x = 0}^{k - 1} \left[ {k \choose x} \left( \frac{2}{3} \right)^x \left( \frac{1 - \gammaB}{6} \right)^{k - x} \theta(x, p^*(x,\cdot)) \right] + \left( \frac{2}{3} \right)^k \theta'(k, p^*(k, \cdot)) \\
= & \sum_{x = 0}^{k} \left[ {k \choose x} \left( \frac{2}{3} \right)^x \left( \frac{1 - \gammaB}{6} \right)^{k - x} \theta(x, p^*(x,\cdot)) \right] - \left( \frac{2}{3} \right)^k \frac{\gammaB}{(1 - \gammaA)(1-\gammaB)} \left( \frac{1 - \gammaA}{2} \right)^k \\
= & \left( \frac{1 - \gammaB}{6} \right)^k \sum_{x = 0}^{k} \left[ {k \choose x} \left( \frac{4}{1 - \gammaB} \right)^x \theta(x, p^*(x,\cdot)) \right] - \frac{\gammaB}{(1 - \gammaA)(1-\gammaB)} \left( \frac{1 - \gammaA}{3} \right)^k \\
= & \left( \frac{1-\gammaB}{6} \right)^k \sum_{x = 0}^k {k \choose x} \left( \frac{4}{1 - \gammaB} \right)^x \left[ c_1 \left( \frac{3 - \gammaB}{4} \right)^x + c_2 \left( \frac{(3 - \gammaB)(1 - \gammaA)}{4} \right)^x \right] \\
& \quad - \left( \frac{1-\gammaB}{6} \right)^k  \frac{\gammaA(1 + \gammaB)^2}{(1-\gammaA) (1 - \gammaB) (3-\gammaB)^2} - \frac{\gammaB}{(1 - \gammaA)(1-\gammaB)} \left( \frac{1 - \gammaA}{3} \right)^k \\
= & c_1 \left( \frac{1 - \gammaB}{6} \right)^k \sum_{x = 0}^k {k \choose x} \left( \frac{3 - \gammaB}{1 - \gammaB} \right)^x + c_2 \left( \frac{1 - \gammaB}{6} \right)^k \sum_{x = 0}^k {k \choose x} \left( \frac{(3 - \gammaB)(1 - \gammaA)}{(1 - \gammaB)} \right)^x\\
& \quad - c_3 t_3^k - c_4 t_4^k\\
= & c_1 \left( \frac{2 - \gammaB}{3} \right)^k + c_2 \left( \frac{4 - 3\gammaA -2\gammaB +\gammaA \gammaB}{6} \right)^k - c_3 t_3^k - c_4 t_4^k\\
= & c_1 t_1^k + c_2 t_2^k - c_3 t_3^k - c_4 t_4^k.
\end{align*}

To see $\eta(k)$ is decreasing over $k$ for $\gammaA = 1/16, \gammaB =  \frac{13 \sqrt{13} - 35}{108}$, we consider the difference between every consecutive terms. Since $\eta(1) = 2/3$, it is easy to observe that $\eta(0) > \eta(1)$. For $k > 0$, we have
\begin{align*}
\eta(k) - \eta(k+1) & = c_1(1-t_1)t_1^k + c_2(1-t_2)t_2^k - c_3(1-t_3)t_3^k - c_4(1-t_4)t_4^k \\
& > (0.3 t_1^k - 0.1 t_4^k) + (0.07 t_2^k - 0.01 t_3^k) > 0,
\end{align*}
where the first inequality derives from
\[
 c_1(1-t_1) > 0.3, \; c_2(1-t_2) > 0.07, \; c_3(1-t_3) < 0.01, \text{ and } c_4(1-t_4) < 0.1,
\]
and the last inequality comes from the fact that $t_1 > t_4$ and $t_2 > t_3$ for the given $\gammaA$ and $\gammaB$.
\end{proof}

We further bound $\eta(k)$ in order to utilize these values in the weighted case. Let $\delta_1 = 0.0309587$ and $\delta_2 = 0.0165525$. Here we note that we obtain these values by rounding off the solution to the system of the following equations:
\[
\left( \frac{2}{3} \right)^2 (1 - \delta_1) = \eta(2) \text{ and } \left( \frac{2}{3} \right)^3 (1 - \delta_1)^2 (1-\delta_2) = \eta(3).
\]
\begin{lemma} \label{lem:weighted-eta}
For $\gammaA = 1/16, \gammaB = \frac{13 \sqrt{13} - 35}{108}$, we have, for every $k \geq 0$,
\[
\eta(k) \leq \left( \frac{2}{3} \right)^k (1 - \delta_1)^{\max(k - 1, 0)} (1 - \delta_2)^{\max(k - 2, 0)}.
\]
\end{lemma}
\begin{proof}
For $k = 0, \cdots, 3$, we have $\eta(k) \leq (2/3)^k (1 - \delta_1)^{\max(k - 1, 0)} (1 - \delta_2)^{\max(k - 2, 0)}$ by construction. For $k = 4, \cdots, 7$, we calculate the following results:
\begin{align*}
& \eta(4) < 0.173 < \left( \frac{2}{3} \right)^4 (1 - \delta_1)^3 (1 - \delta_2)^2; \\
& \eta(5) < 0.11 < \left( \frac{2}{3} \right)^5 (1 - \delta_1)^4 (1 - \delta_2)^3; \\
& \eta(6) < 0.07 < \left( \frac{2}{3} \right)^6 (1 - \delta_1)^5 (1 - \delta_2)^4; \\
& \eta(7) < 0.044 < \left( \frac{2}{3} \right)^7 (1 - \delta_1)^6 (1 - \delta_2)^5,
\end{align*}
satisfying the inequality of the lemma. For $k \geq 8$, let 
\[
d := \frac{1}{(1-\delta_1)(1-\delta_2)^2} \text{ and } s := \left( \frac{2(1-\delta_1)(1-\delta_2)}{3} \right),
\]
and we can observe that
\begin{align*}
\left( \frac{2}{3} \right)^k (1 - \delta_1)^{k - 1} (1 - \delta_2)^{k - 2} - \eta(k) & \geq d \cdot s^k - c_1 \cdot t_1^k - c_2 \cdot t_2^k \\
															& \geq (d - c_1) \cdot s^k - c_2 \cdot t_2^k \\
															& \geq 0.1121 \cdot 0.6353^k - 0.1768 \cdot 0.6^k \\
															& \geq 0,
\end{align*}
where the second inequality comes from the fact that $s > t_1$ and the third inequality follows from
\[
d > 1.0699,\; c_1 < 0.9578,\; c_2 < 0.1768,\; s > 0.6353, \text{ and } t_2 < 0.6.
\]
The last inequality holds for $k \geq 8$.
\end{proof}
\section{General Bound} \label{sect:extend}
In this appendix, we complete the rest of the proof from Section~\ref{sect:general}.

\subparagraph*{Removing Type~\ref{badtype2} Violations.}
We build the new input to $\A$ by inserting $\jbar = (v,\star)$ right before $j$, where $\star$ is a new unique element. Let us fix the first two random bits of every pair (including $\jbar$). We claim that the marginal distributions of the pairs in the subsequences are identical in both inputs.

Consider the ex-post graph of the two inputs. They will be almost identical only with the following possible differences:
\begin{itemize}
\item[-] $(\jhat, j)$ may be only in the original ex-post graph;
\item[-] $(\jhat, \jbar)$ or$ (\jbar, j)$ may be only in the new ex-post graph.
\end{itemize}
For each pair $x$ in the subsequences, the ex-post graph determines, among the third random bits of all pairs, which one determines the output choice of $x$.

Suppose $(\jhat, j)$ is in the original ex-post graph. In this case, $j$ is the only pair in the subsequences whose output choice is determined by the third random bit of $\jhat$. (Recall that $\jhat$ is not in any subsequence.) In the modified input, if $(\jbar, j)$ is in the ex-post graph, $j$ will be the only pair in the subsequences whose output choice is determined by the third random bit of $\jbar$. If $(\jbar, j)$ is not in the ex-post graph, $j$ will be the only pair whose output choice is determined by $j$. Note that the output choice of every other pair in the subsequences is determined by the same random bit in both inputs. Again, since the third random bits of the pairs are i.i.d., this shows that the marginal distribution stays the same.

Suppose now $(\jhat, j)$ is not in the orignal ex-post graph. Let us focus on $j$, since it is the only pair whose output choice may be determined by a different random bit in the modified input. If $j$ was adjacent with another pair in the original ex-post graph, the edge will remain in the new ex-post graph, too, in which case there is nothing to prove. Otherwise, it may be the case that $(\jbar, j)$ may be newly introduced as an edge in the ex-post graph, but this only means that $j$ will be the only pair in the subsequences whose output choice is determined by the third random bit of $\jbar$ (instead of $j$). The marginal distribution therefore stays the same.

\subparagraph*{Further Modification.}
We can remove all Type~\ref{badtype1} and Type~\ref{badtype2} violations after finitely many modifications of the input. We can interpret $\A$ as an algorithm that first constructs the ex-ante graph using the first two random bits and then determines the output using only the third random bits. Our next modification will delete some edges directly from the ex-ante graph rather than modifying the input pairs. (You could alternatively think of this edge being ``disabled,'' which can never appear in the ex-post graph even if it is chosen by the first two random bits of the pairs.) We will now show how we can modify this input graph without affecting \eqref{eq:jointprobobj}.
Suppose we have some pair $j'$ and $j$ such that
\begin{itemize}
\item[-] $j'$ appears before $j$,
\item[-] $j'$ and $j$ are adjacent in the ex-ante graph, and
\item[-] $j$ is not in any subsequence.
\end{itemize}
If we delete this edge from the ex-ante graph, this can change the output choice of only $j$ in $\A$'s output. (Imagine we fix all random bits, and we can easily observe this fact.) Therefore, we can safely delete all such edges without affecting the marginal distribution of the pairs in the subsequences.

Once we remove all such edges, the following argument shows that no two pairs in different subsequences can belong to the same connected component of the ex-ante graph: since two pairs that are not in any subsequences cannot be adjacent, the only way of having two pairs from different subsequences in the same connected component is by having a direct edge in-between (which would be a Type~\ref{badtype1} violation) or a length-2 path whose ``midpoint'' is a predecessor of the two pairs (which would be a Type~\ref{badtype2} violation: note that, in the definition of Type~\ref{badtype2} violation, $\jhat$ cannot be $\pred{j,u}$ because $j'$ appears before $j$).

\subparagraph*{Conclusion.}
Let $p'(\xvect, \yvect)$ be the probability that $\A$ chooses $y_i$ number of $u$'s from the $i$-th subsequence of pairs for all $i$ in the \emph{modified} input. We have so far shown that $p_0 \leq \sum_{\yvect \leq \xvect} p'(\xvect, \yvect) \prod_{i = 1}^m \left( \frac{1}{2} \right)^{k_i - x_i + y_i} (1 - \gammaB)^{\max(k_i - x_i + y_i - 1, 0)}$.
For each $i$, let $p'_i(x_i, y_i)$ be the probability that $\A$ chooses $y_i$ number of $u$'s from the $i$-th subsequence of pairs.
Since $\A$ may introduce negative correlation only on those pair of vertices that are adjacent in the ex-ante graph, output choices made across different connected components of the ex-ante graph will be independent. This implies that $p'(\xvect, \yvect)=\prod_{i=1}^m p'_i(x_i, y_i)$, yielding
\begin{align*}
p_0 &\leq \sum_{\yvect \leq \xvect} p'(\xvect, \yvect) \prod_{i = 1}^m \left( \frac{1}{2} \right)^{k_i - x_i + y_i} (1 - \gammaB)^{\max(k_i - x_i + y_i - 1, 0)}\\
&= \sum_{\yvect \leq \xvect}  \prod_{i = 1}^m p'_i(x_i,y_i) \left( \frac{1}{2} \right)^{k_i - x_i + y_i} (1 - \gammaB)^{\max(k_i - x_i + y_i - 1, 0)}\\
&= \prod_{i = 1}^m \sum_{y_i=0}^{x_i} p'_i(x_i,y_i)   \left( \frac{1}{2} \right)^{k_i - x_i + y_i} (1 - \gammaB)^{\max(k_i - x_i + y_i - 1, 0)}\\
&\leq \prod_{i = 1}^m \sum_{y_i=0}^{x_i} p^*(x_i,y_i)   \left( \frac{1}{2} \right)^{k_i - x_i + y_i} (1 - \gammaB)^{\max(k_i - x_i + y_i - 1, 0)},
\end{align*}
where the last inequality follows from the proof of Lemma~\ref{lem:singleconsecbound}. Since $p^*(x, \cdot)$ depends only on $x$, rather than the actual input, the bound on $p_0$ we obtain depends only on $\xvect$.
Let $\binomprob(\kvect, \xvect, r) := \prod_{i = 1}^m \binomprob(k_i, x_i, r)$. Now the probability that our three-way OCS never chooses $u$ from the given subsequences is no greater than
\begin{align*}
&\sum_{\xvect \leq \kvect} \binomprob \left(\kvect, \xvect, \frac{2}{3} \right) \left[ \prod_{i = 1}^m \sum_{y_i = 0}^{x_i} p^*(x_i, y_i) \left( \frac{1}{2} \right)^{k_i - x_i + y_i} (1 - \gammaB)^{\max(k_i - x_i + y_i - 1, 0)} \right] \\
&= \sum_{\xvect \leq \kvect} \left[ \prod_{i = 1}^m \binomprob \left(k_i, x_i, \frac{2}{3} \right) \right] \left[ \prod_{i = 1}^m \sum_{y_i = 0}^{x_i} p^*(x_i, y_i) \left( \frac{1}{2} \right)^{k_i - x_i + y_i} (1 - \gammaB)^{\max(k_i - x_i + y_i - 1, 0)} \right]\\
&=\prod_{i = 1}^m \sum_{x_i = 0}^{k_i} \binomprob \left(k_i, x_i, \frac{2}{3}\right) \left[ \sum_{y_i = 0}^{x_i} p^*(x_i, y_i) \left( \frac{1}{2} \right)^{k_i - x_i + y_i} (1 - \gammaB)^{\max(k_i - x_i + y_i - 1, 0)} \right]\\
&=\prod_{i = 1}^m \eta(k_i).
\end{align*}
This completes the proof of Theorem~\ref{thm:mainrestate}.

\section{Unweighted Online Bipartite Matching} \label{sect:unweight}
In this appendix, we present how our three-way OCS can be applied to the unweighted
online bipartite matching problem.

Note that the $\frac{13 \sqrt{13} - 35}{108}$-OCS of Fahrbach et al.~\cite{fahrbach} yields a slightly stronger bound than Lemma~\ref{lem:gammaB} when we focus on a single consecutive subsequence.
\begin{lemma}[Fahrbach et al.~\cite{fahrbach}] \label{lem:singlegammaocs}
There exists a two-way OCS such that, for any element $u$ and a consecutive subsequence of pairs containing $u$ of length $k$, the probability that $u$ never gets chosen by the OCS from the given subsequence is at most $\left( \frac{1}{2} \right)^k \cdot f_k$ where $f_k$ is defined recursively as follows:
\[
\textstyle f_k = \left\{
\begin{aligned}
& 1, & \textnormal{if}\ k = 0, 1, \\
&\textstyle   f_{k - 1} - \frac{13 \sqrt{13} - 35}{108} \cdot f_{k - 2}, & \textnormal{otherwise.}
\end{aligned}
\right.
\]
\end{lemma}
 
For each $k = 0, 1, \cdots$, let $\zeta(k)$ be the upper bound of the probability that this two-way OCS never chooses an element from a consecutive subsequence containing the element of length $k$, guaranteed by Lemma~\ref{lem:singlegammaocs}. That is, $\zeta(k) := \left( \frac{1}{2} \right)^k \cdot f_k $. Recall the definition of $\eta(\ell)$ from Section~\ref{ss:etaeta}.

Let $Q := (\Z_+ \times \Z_+ \anhc{) \cup \{(\infty,\infty)\}}$, and sort $Q$ in the ascending order of $1 - \zeta(k) \eta(\ell)$ for each $(k, \ell) \in Q$ (ties are broken arbitrarily). We remark that this value is the lower bound on the probability that an element is chosen at least once when the element is given $k$ times to the two-way OCS and $\ell$ times to our three-way OCS.

\subsection{Algorithm}
For each offline vertex $u$, we maintain $(k, \ell)$ where $k$ and $\ell$ are the number of times that $u$ has been given to the two-way OCS and our three-way OCS, respectively.  Upon each arrival of an online vertex $v$, find adjacent offline vertices $N^*(v) \subseteq N(v)$ \anhc{that come the earliest among $N(v)$ under the ordering given by the sorted $Q$. If every vertex in $N(v)$ was deterministically matched, we leave $v$ exposed.} If $|N^*(v)| = 1$, we deterministically match $v$ to the only vertex in $N^*(v)$. If $|N^*(v)| = 2$, we let the two-way OCS choose an offline vertex and match $v$ to the chosen one. If $|N^*(v)| \geq 3$, we arbitrarily choose three offline vertices from $N^*(v)$, let the three-way OCS choose one vertex, and match $v$ to the chosen one. \anhc{What remains to be defined is the behavior of the algorithm when the chosen offline vertex is already matched. Assume for simplicity that the offline vertex will be re-matched with $v$; however, it is easy to observe that the competitive ratio remains the same even if we leave $v$ exposed in this case.}

\subsection{Primal-Dual Analysis}
We analyze the algorithm for this problem (and the edge-weighted problem) using primal-dual analysis~\cite{devanur13randomized,fahrbach}. Following is an LP relaxation of the unweighted bipartite matching problem:
\begin{align*}
\text{maximize } 	& \sum_{e \in E} x_e & \\
\text{subject to } 	& \sum_{v : v \in N(u)} x_{uv} \leq 1, & \forall u \in L, \\
 				& \sum_{u : u \in N(v)} x_{uv} \leq 1, & \forall v \in R, \\
				& x_e \geq 0, & \forall e \in E,
\end{align*}
where $N(v)$ is the set of vertices adjacent with $v \in L \cup R$. Its dual is as follows:
\begin{align*}
\text{minimize } 	& \sum_{u \in L} \alpha_u + \sum_{v \in R} \beta_v & \\
\text{subject to } 	& \alpha_u + \beta_v \geq 1, & \forall (u, v) \in E, \\
 				& \alpha_u \geq 0, & \forall u \in L, \\
 				& \beta_v \geq 0, & \forall v \in R. 
\end{align*}
According to the standard primal-dual analysis technique~\cite{devanur13randomized,fahrbach}, we will use the following lemma.
\begin{lemma} \label{lem7}
\anhc{Let $\textnormal{\textsf{ALG}}$ denote the number of edges matched by the algorithm.} For some $\Gamma$, if the algorithm can maintain $x$ and $(\alpha, \beta)$ such that 
\begin{enumerate}
\item (Reverse weak duality) $\anhc{ \E[\textnormal{\textsf{ALG}}]\geq } \sum_{e \in E} x_e \geq \sum_{u \in L} \alpha_u + \sum_{v \in R} \beta_v$, and
\item (Approximate dual feasibility) $\alpha_u + \beta_v \geq \Gamma$ for all $(u, v) \in E$,
\end{enumerate}
then the algorithm is $\Gamma$-competitive.
\end{lemma}

\subsection{Primal Variable Construction}
For an offline vertex $u$, let $k$ and $\ell$ denote the number of times $u$ has been given to the two-way OCS and the three-way OCS until the beginning of this iteration.
We aim at setting the primal variables $x$ so as to ensure that
\[
 1 - \zeta(k) \eta(\ell) =  \sum_{v  \in N(u)} x_{(u, v)} =: x_u,
\]
which is a valid lower bound on the probability that $u$ is matched.

Consider a timestep when $v$ arrives. Suppose that $v$ becomes matched during this iteration. If $v$ is matched to $u$ deterministically, we set
\[
x_{(u, v)} = \zeta(k) \eta(\ell),
\]
which in turn makes $x_u = 1$ at the end of this iteration.

On the other hand, if $u_1$ and $u_2$ (with the same $(k, \ell)$) are passed to the two-way OCS in this iteration, we set
\[
x_{(u_1, v)} = x_{(u_2, v)} = \eta(\ell) \left( \zeta(k) - \zeta(k + 1) \right),
\]
resulting in  $x_{u_1} = x_{u_2} = 1 - \zeta(k + 1)\eta(\ell)$ at the end.

Finally, if $u_1$, $u_2$, and $u_3$ (again, with the same $(k, \ell)$) are given to the three-way OCS, we  set
\[
x_{(u_1, v)} = x_{(u_2, v)} = x_{(u_3, v)} = \zeta(k) \left( \eta(\ell) - \eta(\ell + 1) \right).
\]
This again makes $x_{u_1}=x_{u_2}=x_{u_3}= 1 - \zeta(k) \eta(\ell + 1)$ satisfied at the end of the  iteration.

\subsection{Dual Variable Construction}
For the offline vertices, we ensure that the vertices have the same potential if they have been passed to the two-way OCS and the three-way OCS the same number of times. In light of that, we let $\alpha_u$ be $a(k, \ell)$ if $u$ has been input $k$ times to the two-way OCS and $\ell$ times to the three-way OCS at the beginning of each iteration. If $u$ has been deterministically  matched, we let $\alpha_u := a(\infty, \infty)$.

For each $(k, \ell) \in Q$, let $\nextpair{k, \ell}$ be the tuple that immediately follows $(k, \ell)$ in the sorted $Q$. In order to make $a$ monotone, we further ensure that, for each $(k, \ell) \in Q$, we have
\begin{equation} \label{eq1}
a(k, \ell) \leq a(\nextpair{k, \ell}).
\end{equation}
We also require
\begin{equation} \label{eq2}
a(0, 0) = 0.
\end{equation}

For the online vertices, we define $b(k, \ell)$ for each $(k, \ell) \in Q$ and set $\beta_v$ as follows, depending on  how $v$ is matched: If $v$ is matched to an offline vertex with $(k, \ell)$ (at the beginning of this iteration) via the three-way OCS, we set $\beta_v := b(k, \ell)$. On the other hand, if it is matched through the two-way OCS or deterministically, we set $\beta_v := b(\nextpair{k, \ell})$. Finally, if $v$ remains unmatched, we set $\beta_v := 0$.

\subsection{Reverse Weak Duality}
Let us verify the first condition of Lemma~\ref{lem7},  \emph{reverse weak duality}. Since \eqref{eq2} guarantees that the primal and dual objective values are initially the same, it suffices to ensure that, for each iteration, the increment of the primal objective function is always no less than the increment of the dual objective function. 

Fix an iteration. If the online vertex remains exposed, it is trivial since we modify nothing. Now, suppose that the online vertex $v$ is matched to some offline vertex. We have following three cases to handle.

\subparagraph*{Case 1. $v$ is matched deterministically.}
In this case, the primal increase is $ \zeta(k) \eta(\ell) $, while the total dual increment is $a(\infty, \infty) - a(k, \ell) + b(\nextpair{k, \ell})$. Therefore, we need to satisfy, for every  $(k, \ell)$,
\begin{equation} \label{eq3-1}
a(\infty, \infty) - a(k, \ell) + b(\nextpair{k, \ell}) \leq \zeta(k) \eta(\ell).
\end{equation}

\subparagraph*{Case 2. $v$ is matched via the two-way OCS.}
The primal increment is $2  \eta(\ell) \left( \zeta(k) - \zeta(k + 1) \right)$. Each offline vertex passed to the two-way OCS increases its potential by $a(k + 1, \ell) - a(k, \ell)$, and $\beta_v$ is set to be $b(\nextpair{k, \ell})$. We thus have to ensure that, for every  $(k, \ell)$,
\begin{equation} \label{eq3-2}
2 \left( a(k +1, \ell) - a(k, \ell) \right) + b(\nextpair{k, \ell}) \leq 2 \eta(\ell) \left( \zeta(k) - \zeta(k + 1) \right).
\end{equation}

\subparagraph*{Case 3. $v$ is matched via the three-way OCS.}
This case is also similar to the above cases; the primal increment is now $3  \zeta(k) \left( \eta(\ell) - \eta(\ell + 1) \right)$. We increase the dual variables of the offline vertices passed to the three-way OCS by $a(k, \ell + 1) - a(k, \ell)$ each, and we set the dual variable of $j$ as $b(k, \ell)$. Therefore, we have to ensure that, for every $(k, \ell)$,
\begin{equation} \label{eq3-3}
3 \left( a(k, \ell + 1) - a(k, \ell) \right) + b(k, \ell) \leq 3 \zeta(k) \left( \eta(\ell) - \eta(\ell + 1) \right).
\end{equation}

\subsection{Approximate Dual Feasibility}
Now we verify the second condition,  \emph{approximate dual feasibility}. For each $(u, v) \in E$, it suffices to ensure that this condition holds when $v$ arrives because $\alpha_u$ only increases over iterations and $\beta_v$ never changes once it is set.

If $v$ remains exposed, this means that every adjacent offline vertex is already deterministically matched. Therefore, we should guarantee
\begin{equation} \label{eq4-1}
a(\infty, \infty) \geq \Gamma.
\end{equation}

Otherwise, note that $N^*(v)$ is the set of the adjacent offline vertices with the most preceding $(k, \ell)$ and we ensure monotonically increasing $a$ in the order of the sorted $Q$ due to~\eqref{eq1}. Moreover, if $v$ is matched to an offline vertex with $(k, \ell)$ (at the beginning) deterministically or via the two-way OCS, every vertex adjacent with $v$ will have $(k', \ell')$ that is strictly after $(k, \ell)$ in the sorted $Q$ by the end of the iteration. This justifies our choice of $\beta_v$. Therefore, it suffices to have, for every $(k, \ell)$,
\begin{equation} \label{eq4-2}
a(k, \ell) + b(k, \ell) \geq \Gamma.
\end{equation}

\subsection{A Factor-Revealing LP}\label{ss:ufrl}
We can now obtain the algorithm's performance guarantee by solving a factor-revealing LP \anhc{that determines the values of $a$ and $b$ while} ensuring the constraints \anhc{derived} in the previous appendices. \anhc{Let us use $\leq_Q$ to denote the ordering given by the sorted $Q$: if $(k,\ell)$ precedes or is equal to $(k',\ell')$, we say $(k,\ell)\leq_Q (k',\ell')$. Otherwise, we say $(k,\ell)>_Q (k',\ell')$.

We need to have a finite LP to computationally solve it. We will choose some parameter $(\kmax, \ellmax)$ and consider $a(k,\ell)$ and $b(k,\ell)$ only for $(k,l)\leq_Q (\kmax, \ellmax)$. This keeps the number of LP variables finite.
For $(k,l)>_Q (\kmax, \ellmax)$, we define that $a(k,l):=a(\kmax,\ellmax)$ and $b(k,l):=0$. This choice lets us keep the number of constraints finite as well.
(Technically, we will create LP variables $a(k,l)$ and $b(k,l)$ for a few additional $(k,l)$ that follows $(\kmax, \ellmax)$ too, because they appear on some constraints. However, their values will be constrained as described above: $a(k,l)=a(\kmax,\ellmax)$ and $b(k,l)=0$.)
}
\begin{lemma}
\anhc{The competitive ratio of the algorithm is greater than or equal to the value of the following LP.}
\begin{align}
\textnormal{maximize } & \Gamma & \nonumber\\
\textnormal{subject to } & a(k, \ell) \geq 0, & \forall \anhc{(k, \ell) \leq_Q (\kmax, \ellmax)}, \label{fl9}\\
 & b(k, \ell) \geq 0, & \forall \anhc{(k, \ell) \leq_Q (\kmax, \ellmax)},\label{fl10}\\
& a(0, 0) = 0, & \label{fl0}\\
 & a(k, \ell) \leq a(\nextpair{k, \ell}), & \forall (k, \ell) \leq_Q (\kmax, \ellmax), \label{fl1}\\
 & a(\kmax, \ellmax) - a(k, \ell) + b(\nextpair{k, \ell}) &\nonumber \\
 & \hspace{7em} \leq \zeta(k) \eta(\ell), & \forall (k, \ell) \leq_Q (\kmax, \ellmax), \label{fl2}\\
 & 2 \left( a(k + 1, \ell) - a(k, \ell) \right) + b(\nextpair{k, \ell}) &\nonumber\\
 & \hspace{7em} \leq 2\eta(\ell)\left( \zeta(k) - \zeta(k + 1) \right), & \forall (k, \ell) \leq_Q (\kmax, \ellmax), \label{fl3}\\
 & 3 \left( a(k, \ell + 1) - a(k, \ell) \right) + b(k, \ell) &\nonumber\\
 & \hspace{7em} \leq 3\zeta(k)\left( \eta(\ell) - \eta(\ell + 1) \right), & \forall (k, \ell) \leq_Q (\kmax, \ellmax), \label{fl4}\\
 & a(\kmax, \ellmax) \geq \Gamma, & \label{fl5}\\
 & a(k, \ell) + b(k, \ell) \geq \Gamma, & \forall (k, \ell) \leq_Q (\kmax, \ellmax), \label{fl6}\\
 & \anhc{a(k, \ell) = a(\kmax, \ellmax),} & \anhc{\forall (k,\ell) >_Q (\kmax, \ellmax),}\label{fl7}\\
 & \anhc{b(k, \ell) = 0,} & \anhc{\forall (k,\ell) >_Q (\kmax, \ellmax).}\label{fl8}
\end{align}
\end{lemma}
\anhc{In actual computation, we create only those variables required by Constraints~\eqref{fl1}-\eqref{fl6}. Note that it is sufficient to impose the Constraints~\eqref{fl2}-\eqref{fl4} only for $(k, \ell) \leq_Q (\kmax, \ellmax)$ because the constraints corresponding to $(k,\ell) >_Q (\kmax, \ellmax)$ will be trivially satisfied by our choice of $a(k,\ell)$ and $b(k,\ell)$ for $(k,\ell) >_Q (\kmax, \ellmax)$. Similarly, Constraint~\eqref{fl6} for $(k,\ell) >_Q (\kmax, \ellmax)$ are satisfied due to Constraint~\eqref{fl5}.}

With $(\kmax, \ellmax) = (8, 0)$, we obtain $\Gamma = 0.50962346$. This completes the proof of Theorem~\ref{thm:uwm}.

\section{Edge-Weighted Online Bipartite Matching} \label{sect:weight}
In this appendix, we present how we can apply our three-way OCS to  edge-weighted online bipartite matching with free disposal. In particular, we give a $0.5093$-competitive algorithm for the problem. We will slightly abuse the notation and redefine $\zeta$ and $\eta$ in this appendix so that they denote the slightly weaker bounds from Lemmas~\ref{lem:gammaB} and \ref{lem:weighted-eta} (and Theorem~\ref{thm:mainrestate}): i.e.,
\begin{eqnarray*}
\zeta(k) &:=& \left( \frac{1}{2} \right)^k (1 - \gamma)^{\max(k - 1, 0)}\ \textnormal{and}\\
\eta(\ell) &:=& \left( \frac{2}{3} \right)^\ell (1 - \delta_1)^{\max(\ell - 1, 0)} (1 - \delta_2)^{\max(\ell - 2, 0)},
\end{eqnarray*}
where $\gamma := \frac{13 \sqrt{13} - 35}{108}$.

\subsection{Algorithm}
Let $a : \Z_+ \times \Z_+ \cup \{ (\infty, \infty) \} \to \R_+$ and $b : \Z_+ \times \Z_+ \cup \{ (\infty, \infty) \} \to \R_+$ be two functions whose values will be defined later by a factor-revealing LP. Let $\sigRtwo, \sigD \in \R_+$ be two parameters satisfying
\begin{equation} \label{weighted:eq-sigmas}
0 < \sigRtwo \leq \frac{3}{2} \text{ and } 0 < \sigD \leq \frac{3 \sigRtwo}{3 - \sigRtwo}.
\end{equation}
Later we will set $\sigRtwo = 1.3$ and $\sigD = 2.2$.

The overall structure of the algorithm is as follows. The algorithm internally runs the two-way $\gamma$-OCS from Lemma~\ref{lem:gammaB} and our three-way OCS. Upon the arrival of an online vertex $v$, it performs one of the following:\begin{itemize}
\item selects a triple of three offline vertices, passes it to the three-way OCS, and matches $v$ with the output of the three-way OCS;
\item selects a pair of two offline vertices, passes it to the two-way OCS, and matches $v$ with the output of the two-way OCS;
\item selects one offline vertex, and deterministically matches $v$ with the selected offline vertex; or
\item leaves $v$ exposed.
\end{itemize}

Let $u\in L$ be an offline vertex. If the algorithm decided to pass a triple containing $u$ to the three-way OCS when an online vertex $v\in R$ arrived, we define the \emph{weight level} of that triple as the weight of the edge $(u,v)$. Note that this definition is dependent on $u$: the triple passed to the three-way OCS at this timestep is considered to have three different weight levels, depending on which offline vertex we are interested in.
Similarly, if the algorithm passed a pair containing $u$ when $v$ arrived, the weight level of that pair is $w_{uv}$.

Throughout the algorithm, for each offline vertex $u \in L$ and every weight level $w \in \R_+$, we maintain the number of times that $u$ has been passed to the two-way OCS in a pair of weight level at least $w$, denoted by $k_u(w)$, and the number of times $u$ has been passed to the three-way OCS in a triple of weight level at least $w$, denoted by $\ell_u(w)$. If $u$ has been deterministically matched with an edge of weight at least $w$, we define $k_u(w) = \ell_u(w) = \infty$.

When an online vertex $v \in R$ arrives, we compute, for every offline vertex $u \in L$,
\begin{align*}
\delRthreeBeta{u}{v} &:= \int^{w_{uv}}_0 b(k_u(w), \ell_u(w)) dw - \frac{1}{3} \int_{w_{uv}}^\infty a(k_u(w), \ell_u(w)) dw, \\
\delRtwoBeta{u}{v} &:= \sigRtwo \delRthreeBeta{u}{v}, \text{ and} \\
\delDBeta{u}{v} &:=  \sigD \delRthreeBeta{u}{v}.
\end{align*}
We then select three offline vertices achieving the maximum value of $\delRthreeBeta{u}{v}$, denoted by $u_1$, $u_2$, and $u_3$; we also select two vertices with maximum $\delRtwoBeta{u}{v}$, denoted by $\ubar_1$ and $\ubar_2$, and the vertex $\ustar$ with maximum $\delDBeta{u}{v}$. For notational simplicity, we let
\[
\betaRthree := \delRthreeBeta{u_1}{v} + \delRthreeBeta{u_2}{v} + \delRthreeBeta{u_3}{v},\; \betaRtwo := \delRtwoBeta{\ubar_1}{v} + \delRtwoBeta{\ubar_2}{v}, \text{ and } \betaD := \delDBeta{\ustar}{v}.
\]

If none of $\betaRthree$, $\betaRtwo$, and $\betaD$ is greater than $0$, $v$ remains exposed. Otherwise, we find the largest value among $\betaRthree$, $\betaRtwo$, and $\betaD$. If $\betaRthree$ is the largest, we let the three-way OCS choose among $u_1$, $u_2$, and $u_3$. If $\betaRtwo$ is the largest, we send $\ubar_1$ and $\ubar_2$ to the two-way OCS. Finally, if $\betaD$ is the largest, we deterministically choose $\ustar$. After matching $v$ with the chosen vertex, we update $k_u(w)$ and $\ell_u(w)$ accordingly.

\subsection{Primal-Dual Analysis}
We use the weighted version of Lemma~\ref{lem7}, stated below.
\begin{restatable}{lemma}{primaldual} \label{lem:primaldual}
Let $\textnormal{\textsf{ALG}}$ be the total weight of the edges matched by an algorithm. For some $\Gamma$, if the algorithm always maintains $(\alpha, \beta)$ satisfying
\begin{enumerate}
\item (Reverse weak duality) $\E[\textnormal{\textsf{ALG}}] \geq \sum_{u \in L} \alpha_u + \sum_{v \in R} \beta_v$, and
\item (Approximate dual feasibility) for each $(u, v) \in E$, $\alpha_u + \beta_v \geq \Gamma w_{uv}$,
\end{enumerate}
then the algorithm is $\Gamma$-competitive.
\end{restatable}

\subsection{Primal Variable Construction}
Instead of constructing primal variables $x$, we will directly give a bound on the expected weight of the algorithm's output,  adopting a similar approach to~\cite{devanur2016whole, fahrbach}. For each offline vertex $u \in L$ and for every weight level $w \in \R_+$, let $y_u(w)$ be the probability that $u$ is matched with an edge of weight at least $w$. We can then write the expected weight of the edge with which $u$ is matched by
\[
\int_0^{\infty} y_u(w) dw.
\]

Let us fix an offline vertex $u$ and a weight level $w$.
Among all the pairs that have been given to the two-way OCS, if
 we focus on the pairs containing $u$ with weight level at least $w$,  they form a set of disjoint consecutive subsequences of pairs containing $u$. Let $k_1, \cdots, k_m$ be the lengths of these subsequences. Similarly, for the triples containing $u$ with weight level at least $w$, we can view them as a set of disjoint consecutive subsequences of triples of lengths $\ell_1, \cdots, \ell_n$. Observe that the probability $y_u(w)$ that $u$ is matched with an edge of weight at least $w$ is at least
\[
 \ybar_u(w) := 1 - \prod_{i = 1}^m \zeta(k_i) \prod_{j = 1}^m \eta(\ell_i).
\]
This implies that we can obtain a valid lower bound on the expected weight of the matching returned by our algorithm as follows:
\[
P := \sum_{u \in L} \int_0^{\infty} \ybar_u(w) dw.
\]

Here we present several observations that helps us ensure the reverse weak duality in Section~\ref{ss:rwd}.
\begin{observation} \label{weighted:obs1}
For any set of disjoint consecutive subsequences of pairs of lengths $k_1, \cdots, k_m$ passed to the two-way OCS and any set of disjoint consecutive subsequences of triples of lengths $\ell_1, \cdots, \ell_n$ passed to the three-way OCS, we have
\[
\prod_{i = 1}^m \zeta(k_i) \prod_{j = 1}^n \eta(\ell_j) \geq \zeta(k) \eta(\ell),
\]
where $k = k_1 + \cdots + k_m$ and $\ell = \ell_1 + \cdots + \ell_n$.
\end{observation}

\begin{observation} \label{weighted:obs2}
If the algorithm deterministically matches an offline vertex $u$ by an edge of weight at least $w$, then the increment of $\ybar_u(w)$ is at least $\zeta(k_u(w)) \eta(\ell_u(w))$.
\end{observation}

\begin{observation} \label{weighted:obs3}
If the algorithm sends $u$ to the two-way OCS with weight level at least $w$, then the increment of $\ybar_u(w)$ is at least
\[
\frac{1}{2} \zeta(k_u(w)) \eta(\ell_u(w)).
\]
Furthermore, let $w'$ be the weight level of the last pair containing $u$ before this iteration. If $w' \geq w$, then the increment of $\ybar_u(w)$ is at least
\[
\frac{1 + \gamma}{2} \zeta(k_u(w)) \eta(\ell_u(w)).
\]
\end{observation}

\begin{observation} \label{weighted:obs4}
If the algorithm inserts $u$ into the three-way OCS with weight level at least $w$, then the increment of $\ybar_u(w)$ is at least
\[
\frac{1}{3} \zeta(k_u(w)) \eta(\ell_u(w)).
\]
Moreover, let $w'$ be the weight level of the last pair containing $u$ before this iteration. If $w' \geq w$, the increment of $\ybar_u(w)$ is at least
\[
\frac{1 + 2 \delta_1}{3} \zeta(k_u(w)) \eta(\ell_u(w)).
\]
Finally, let $w''$ be the weight level of the second-to-last pair containing $u$. If $\min(w', w') \geq w$, the increment of $\ybar_u(w)$ is at least
\[
\frac{1 + 2 \delta_1 + 2 \delta_2 - 2\delta_1 \delta_2}{3} \zeta(k_u(w)) \eta(\ell_u(w)).
\]
\end{observation}

\subsection{Dual Variable Construction}
Now we describe construction of the dual variables. The dual variable of each online vertex is determined by which method it is matched when it arrives. When an online vertex $v$ arrives, if $v$ is matched through the three-way OCS, then $\beta_v := \betaRthree$. If $v$ is matched through the two-way OCS, $\beta_v := \betaRtwo$. If the algorithm deterministically matches $v$, $\beta_v := \betaD$. Finally, if $v$ remains exposed, we simply set $\beta_v := 0$.

On the other hand, the dual variable of an offline vertex only depends on how many times the vertex gets inserted into the respective OCSs. In particular, for every $u \in L$, we define
\[
\alpha_u := \int_0^\infty \alpha_u(w) dw,
\]
where we will define $\alpha_u(w)$ so that the following invariant is  maintained: for every weight level $w$,
\begin{equation} \label{eq:offlineinvariant}
\alpha_u(w) \geq a(k_u(w), \ell_u(w)).
\end{equation}
Initially, we let $\alpha_u(w) := 0$ for all $u$ and $w$.

Let us exhibit how we update $\alpha_u(w)$ according to the way an offline vertex $v$ gets matched at each iteration. Remark that $k_u(w)$ and $\ell_u(w)$ are the values at the beginning of the iteration (before matching $v$).

\subparagraph*{Matching Deterministically.}
Let us first consider the case when the algorithm deterministically matches $v$ to $u$. By definition, to have the invariant satisfied, we set $\alpha_u(w) = a(\infty, \infty)$ for all  $w \leq w_{uv}$. That is, the increment of $\alpha_u(w)$ at this weight level will be at most
\[
a(\infty, \infty) - a(k_u(w), \ell_u(w)).
\]
For $w > w_{uv}$, $\alpha_u(w)$ remains the same.

\subparagraph*{Matching via the Two-Way OCS.}
Now we suppose that the algorithm passes $u$ to the two-way OCS. Let $w'$ be the weight level of the last pair containing $u$ before this iteration; if $k_u(w) = 0$, let $w' = 0$. We then have the following update rule:

\[
\Delta \alpha_u(w) := \left\{
\begin{aligned}
& a(k_u(w) + 1, \ell_u(w)) - a(k_u(w), \ell_u(w)),	 					&& \text{ if } w \leq \min(w', w_{uv}),\\
& a(k_u(w) + 1, \ell_u(w)) - a(k_u(w), \ell_u(w)),	 					&& \text{ if } w' < w \leq w_{uv} \text{ and } k_u(w) = 0,\\
& a(k_u(w) + 1, \ell_u(w)) - a(k_u(w), \ell_u(w))	 					&& \\
& \hspace{4em} - \frac{\gamma}{2}  \cdot \zeta(k_u(w)) \eta(\ell_u(w)), 	&& \text{ if } w' < w \leq w_{uv} \text{ and } k_u(w) \geq 1, \\
& \frac{\gamma}{2} \cdot \zeta(k_u(w)) \eta(\ell_u(w)),				&& \text{ if } w > w_{uv} \text{ and } k_u(w) \geq 1.
\end{aligned}
\right.
\]
We claim that we still maintain the invariant after the update. Indeed, the only concern is when $w' < w \leq w_{uv}$ and $k_u(w) \geq 1$ since we may have a deficit of at most $\frac{\gamma}{2} \cdot \zeta(k_u(w)) \eta(\ell_u(w))$. However, this deficit can be compensated by the ``prepayment'' made when the last pair was processed: for $w > w'$ and $k_u(w) \geq 1$, we have already paid $\frac{\gamma}{2} \cdot \zeta(k_u(w)) \eta(\ell_u(w))$. Observe that, since the prepayment, $k_u(w)$ cannot have changed until this iteration. It might be the case that $\ell_u(w)$ increased, but it would only make $\frac{\gamma}{2} \cdot \zeta(k_u(w)) \eta(\ell_u(w))$ smaller at this iteration. 

\subparagraph*{Matching via the Three-Way OCS.}
Finally, let us assume that $u$ gets inserted into the three-way OCS.
Let $w'$  be the weight level of the last triple containing $u$ before this iteration; if $\ell_u(w) = 0$, let $w' = 0$. Similarly, let $w''$ be the weight level of the second-to-last triple containing $u$; if $\ell_u(w) \leq 1$, let $w'' = 0$. 
The update rule will again have deficit in some cases which will be compensated by prepayments in a similar manner to the two-way case.
However, there also is a difference: in the two-way case, deficit was always compensated by the prepayment of the last pair; in the three-way case, deficit will be compensated by either the last or the second-to-last triple. As such, the prepayments will always be the sum of two terms: one is to be used by the next triple (if needed) and the other is to be used by the second last triple (if needed).

We will present the update rule broken into three cases; but, the ``prepayment'' rule when $w>w_{uv}$ is common to the three cases and we will present it first.
\[
\Delta \alpha_u(w) := \left\{
\begin{aligned}
& \frac{2 \delta_1}{3} \cdot \zeta(k_u(w)) \eta(\ell_u(w)) && \\
& \hspace{2em} + \frac{2(\delta_2 - \delta_1 \delta_2)}{3} \cdot \zeta(k_u(w)) \eta(\ell_u(w) + 1), && \text{ if } w > w_{uv} \text{ and } \ell_u(w) = 1,\\
& \frac{2 (\delta_1 + \delta_2 - \delta_1 \delta_2)}{3} \cdot \zeta(k_u(w)) \eta(\ell_u(w)) && \\
& \hspace{2em} + \frac{2(\delta_2 - \delta_1 \delta_2)}{3} \cdot \zeta(k_u(w)) \eta(\ell_u(w) + 1), && \text{ if } w > w_{uv} \text{ and } \ell_u(w) \geq 2.
\end{aligned}
\right.\]

Now we present the update rule when $w\leq w_{uv}$, broken into three cases.
The first case is when $w_{uv}$ is the smallest among $w_{uv}, w', w''$:
\[
\Delta \alpha_u(w) := 
 a(k_u(w), \ell_u(w) + 1) - a(k_u(w), \ell_u(w)).
\]
Note that the condition $w\leq w_{uv}$ is implicit above.
We can immediately see that the invariant is satisifed, without the need of compensation.

The second case is when $w'$ is the smallest among $w_{uv}, w', w''$:
\[
\Delta \alpha_u(w) := \left\{
\begin{aligned}
& a(k_u(w), \ell_u(w) + 1) - a(k_u(w), \ell_u(w)), && \text{ if } w \leq w',\\
& a(k_u(w), \ell_u(w) + 1) - a(k_u(w), \ell_u(w)), && \text{ if } w' < w \leq w_{uv} \text{ and } \ell_u(w) = 0, \\
& a(k_u(w), \ell_u(w) + 1) - a(k_u(w), \ell_u(w)) && \\
& \hspace{2em} - \frac{2\delta_1}{3} \cdot \zeta(k_u(w)) \eta(\ell_u(w)), && \text{ if } w' < w \leq w_{uv} \text{ and } \ell_u(w) = 1, \\
& a(k_u(w), \ell_u(w) + 1) - a(k_u(w), \ell_u(w)) && \\
& \hspace{2em} - \frac{2(\delta_1 + \delta_2 - \delta_1 \delta_2)}{3} \cdot \zeta(k_u(w)) \eta(\ell_u(w)), && \text{ if } w' < w \leq w_{uv} \text{ and } \ell_u(w) \geq 2.
\end{aligned}
\right.
\]
In the first two cases, the invariant is trivially maintained.
In the last two cases, we will use the prepayment of $w'$ to compensate for the deficit. Since $w'<w$, the last triple did make the prepayment at weight level $w$. Note that the deficit of these two cases exactly correspond to the first terms of the two cases in the prepayment rule above. It may be the case that $k_u(w)$ increased between the prepayment and the current iteration; however, this is not a problem because it only makes the deficit smaller than the prepayment.

The last case is when $w''$ is the smallest among $w_{uv}, w', w''$:
\[
\Delta \alpha_u(w) := \left\{
\begin{aligned}
& a(k_u(w), \ell_u(w) + 1) - a(k_u(w), \ell_u(w)), && \text{ if } w \leq w'',\\
& a(k_u(w), \ell_u(w) + 1) - a(k_u(w), \ell_u(w)), && \text{ if } w'' < w \leq  \min(w', w_{uv}) \text{ \& } \ell_u(w) \leq 1, \\
& a(k_u(w), \ell_u(w) + 1) - a(k_u(w), \ell_u(w)) && \\
& \hspace{2em} - \frac{2(\delta_2 - \delta_1 \delta_2)}{3} \cdot \zeta(k_u(w)) \eta(\ell_u(w)), && \text{ if } w'' < w \leq  \min(w', w_{uv}) \text{ \& } \ell_u(w) \geq 2, \\
& a(k_u(w), \ell_u(w) + 1) - a(k_u(w), \ell_u(w)), && \text{ if } \min(w', w_{uv}) < w \leq w_{uv} \text{ \& } \ell_u(w) = 0, \\
& a(k_u(w), \ell_u(w) + 1) - a(k_u(w), \ell_u(w)) && \\
& \hspace{2em} - \frac{2\delta_1}{3} \cdot \zeta(k_u(w)) \eta(\ell_u(w)), && \text{ if } \min(w', w_{uv}) < w \leq w_{uv} \text{ \& } \ell_u(w) = 1, \\
& a(k_u(w), \ell_u(w) + 1) - a(k_u(w), \ell_u(w)) && \\
& \hspace{1em} - \frac{2(\delta_1 + \delta_2 - \delta_1 \delta_2)}{3} \cdot \zeta(k_u(w)) \eta(\ell_u(w)), && \text{ if } \min(w', w_{uv}) < w \leq w_{uv} \text{ \& } \ell_u(w) \geq 2. \\
\end{aligned}
\right.
\]
In the first, second, and fourth cases, the invariant is trivially satisfied.
In the fifth and sixth case, we have $w'<w\leq w_{uv}$ and therefore the last triple did make the prepayment that can be used to compensate for the deficit.
In the remaining third case, we have $w''<w$ and therefore the second-to-last triple (which exists because $\ell_u(w)\geq 2$) did make the prepayment. We can see from the prepayment rule that the prepayment allocated for the second next triple is always 
``$\frac{2(\delta_2 - \delta_1 \delta_2)}{3} \cdot \zeta(k_u(w)) \eta(\ell_u(w) + 1)$'', where $k_u(w)$ and $\ell_u(w) $ denote their respective values at the beginning of the iteration when the second-to-last triple arrived. 
In that iteration, $k_u(w)$ was no greater than its current value, and $\ell_u(w)+1$ is equal to the current value of $\ell_u(w)$. This is because $\ell_u(w)$ increased when the last triple was processed. Note that $w\leq w'$ in this case.

\subsection{Reverse Weak Duality}\label{ss:rwd}
At the end of the analysis, we will construct a factor-revealing LP to determine $a$ and $b$, and bound the competitive ratio. In this section, we will derive the constraints that need to be satisfied by  $a$ and $b$ in order to ensure the reverse weak duality of Lemma~\ref{lem:primaldual}. Note that it suffices to ensure that the increase in $P$ is no less than the increase in the dual objective for each iteration, since both $P$ and the dual objective are zero at the beginning of the algorithm.

Let $v\in R$ be the online vertex that arrives at an iteration. If the  algorithm decides to leave $v$ exposed, there is nothing to prove: neither $P$ nor the dual objective changes.

\subparagraph*{Initialization.}
In order to satisfy \eqref{eq:offlineinvariant} at the very beginning of the algorithm, we need
\begin{equation} \label{weighted:const0}
a(0, 0) = 0.
\end{equation}

\subparagraph*{Matching Deterministically.}
By Observation~\ref{weighted:obs2} and the update rule of $(\alpha, \beta)$, it suffices to ensure
\begin{align*}
& \int_0^{w_{\ustar v}} a(\infty, \infty) - a(k_{\ustar}(w), \ell_{\ustar}(w)) dw + \sigD \int_0^{w_{\ustar v}} b(k_{\ustar}(w), \ell_{\ustar}(w)) dw \\
& \hspace{4em} \leq \int_0^{w_{\ustar v}} \zeta(k_{\ustar}(w)) \eta(\ell_{\ustar}(w)) dw,
\end{align*}
where we drop the negative term from $\delDBeta{\ustar}{v}$. We satisfy the above inequality by enforcing a stronger condition: for every $k$ and $\ell$,
\begin{equation} \label{weighted:const1}
a(\infty, \infty) - a(k, \ell) + \sigD \cdot b(k, \ell) \leq \zeta(k) \eta(\ell).
\end{equation}

\subparagraph*{Matching via the Two-Way OCS.}
It is sufficient to have, for each $u = \ubar_1, \ubar_2$, the increment of $P$ due to $\ybar_u$ is no less than the increments of $\alpha_u$ and $\delRtwoBeta{u}{v}$.
We will further ensure that, for all weight level $w \leq w_{uv}$, the increase in $\ybar_u(w)$ is greater than or equal to $\Delta \alpha_u(w) + b(k_u(w), \ell_u(w))$.
Let $w'$ be the weight level of the last pair containing $u$; if $u$ has never appeared in the two-way OCS before, let $w' = 0$.

Let us first consider the weight level $w$ such that $k_u(w) = 0$. By Observation~\ref{weighted:obs3} and the update rule of $(\alpha, \beta)$, it suffices to have
\[
a(1, \ell_u(w)) - a(0, \ell_u(w)) + \sigRtwo b(0, \ell_u(w))  \leq \frac{1}{2} \cdot \zeta(0) \eta(\ell_u(w)).
\]
We satisfy this inequality by the set of following constraints: for every $\ell$,
\begin{equation} \label{weighted:const21}
a(1, \ell) - a(0, \ell) + \sigRtwo b(0, \ell)  \leq \frac{1}{2} \cdot \zeta(0) \eta(\ell) = \frac{\eta(\ell)}{2}.
\end{equation}

Now we consider the case of  $k_u(w) \geq 1$. For $w \leq \min(w', w_{uv})$, by Observation~\ref{weighted:obs3} and the update rule of the dual variables, it would be enough to have
\[
a(k_u(w) + 1, \ell_u(w)) - a(k_u(w), \ell_u(w)) + \sigRtwo b(k_u(w), \ell_u(w))  \leq \frac{1 + \gamma}{2} \cdot \zeta(k_u(w)) \eta(\ell_u(w)),
\]
which will be guaranteed by the following constraints: for every $k \geq 1, \ell$,
\begin{equation} \label{weighted:const22}
a(k + 1, \ell) - a(k, \ell) + \sigRtwo b(k, \ell) \leq \frac{1 + \gamma}{2} \cdot \zeta(k) \eta(\ell).
\end{equation}
In contrast, for $w' < w \leq w_{uv}$, we wrote a different sufficient inequality due to the deficit:
\begin{align*}
& a(k_u(w) + 1, \ell_u(w)) - a(k_u(w), \ell_u(w)) - \frac{\gamma}{2}\zeta(k_u(w)) \eta(\ell_u(w)) + \sigRtwo b(k_u(w), \ell_u(w)) \\
& \hspace{4em} \leq \frac{1}{2} \cdot \zeta(k_u(w)) \eta(\ell_u(w)).
\end{align*}
Note that it is implied by Constraint~\eqref{weighted:const22}.

For $w > w_{uv}$ where $k_u(w) \geq 1$, $\ybar_u(w)$ remains the same. Therefore, the increment in $\Delta \alpha_u(w)$ should be properly covered by the negative term of $\delRtwoBeta{u}{v}$, i.e., we should satisfy:
\[
\frac{\sigRtwo}{3} a(k_u(w), \ell_u(w)) \geq \frac{\gamma}{2} \cdot \zeta(k_u(w)) \eta(\ell_u(w)).
\]
We will later ensure that $a(k, \ell)$ is increasing over $k$ and \anhc{over} $\ell$ while the right-hand side becomes smaller when $k_u(w)$ and $\ell_u(w)$ get larger. Therefore, it is sufficient to guarantee
\begin{equation} \label{weighted:const23}
a(1, 0) \geq \frac{3 \gamma }{2 \sigRtwo} \cdot \zeta(1) \eta(0) = \frac{3 \gamma}{4 \sigRtwo}.
\end{equation}

\subparagraph*{Matching via the Three-Way OCS.}
In this case, we again appeal to a similar argument to the above. For each $u = u_1, u_2, u_3$, we will make sure that the increment of $P$ due to $u$ is always at least the increments of $\alpha_u$ and $\delRthreeBeta{u}{v}$. Here, let $w'$ and $w''$ be the weight level of the last and second-to-last triples containing $u$; if such a triple does not exist, we set $w'$ or $w''$ to be $0$.

For weight level $w$ such that $\ell_u(w) = 0$, due to Observation~\ref{weighted:obs4} and the update rule of the dual variables, it suffices to satisfy
\[
a(k_u(w), 1) - a(k_u(w), 0) + b(k_u(w), 0) \leq \frac{1}{3} \cdot \zeta(k_u(w)) \eta(0),
\]
which will be ensured by the following constraint: for every $k$,
\begin{equation} \label{weighted:const31}
a(k, 1) - a(k, 0) + b(k, 0) \leq \frac{1}{3} \cdot \zeta(k) \eta(0) = \frac{\zeta(k)}{3}.
\end{equation}

Next, for weight level $w$ where $\ell_u(w) = 1$, due to the prepay-compensate technique of the update rule, it suffices to ensure that
\[
a(k_u(w), 2) - a(k_u(w), 1) + b(k_u(w), 1) \leq \frac{1 + 2 \delta_1}{3} \cdot \zeta(k_u(w)) \eta(1),
\]
which will be guaranteed by the following constraint: for every $k$,
\begin{equation} \label{weighted:const32}
a(k, 2) - a(k, 1) + b(k, 1) \leq \frac{1 + 2 \delta_1}{3} \cdot \zeta(k) \eta(1) = \frac{2 + 4 \delta_1}{9} \cdot \zeta(k).
\end{equation}

For weight level $w$ where $\ell_u(w) \geq 2$, it is sufficient for us to obtain
\begin{align*}
& a(k_u(w), \ell_u(w) + 1) - a(k_u(w), \ell_u(w)) + b(k_u(w), \ell_u(w)) \\
& \hspace{8em} \leq \frac{1 + 2 \delta_1 + 2 \delta_2 - 2 \delta_1 \delta_2}{3} \cdot \zeta(k_u(w)) \eta(\ell_u(w)),
\end{align*}
again due to the prepay-compensate technique. We will create constraints in the factor-revealing LP as follows: for each $k$ and $\ell \geq 2$,
\begin{equation} \label{weighted:const33}
a(k, \ell + 1) - a(k, \ell) + b(k, \ell) \leq \frac{1 + 2 \delta_1 + 2\delta_2 - 2\delta_1 \delta_2}{3} \cdot \zeta(k) \eta(\ell),
\end{equation}
which is again adequate for this case.

Finally, we have to bound the prepayment. For weight level $w$ with $\ell_u(w) = 1$, we should guarantee
\[
\frac{2 \delta_1}{3} \cdot \zeta(k_u(w)) \eta(1) + \frac{2 \delta_2 - 2 \delta_1 \delta_2}{3} \cdot \zeta(k_u(w)) \eta(2) - \frac{1}{3} a(k_u(w), 1) \leq 0.
\]
We will later ensure that $a(k, \ell)$ increases over $k$ and $\ell$. Then, since $\zeta(k)$ is decreasing over $k$, it suffices to have
\begin{equation} \label{weighted:const34}
a(0, 1) \geq 2 \delta_1 \cdot \zeta(0) \eta(1) + (2 \delta_2 - 2 \delta_1 \delta_2) \cdot \eta(2).
\end{equation}
On the other hand, for weight level $w$ where $\ell_u(w) \geq 2$, we need to obtain
\begin{align*}
& \frac{2 \delta_1 + 2 \delta_2 - 2 \delta_1 \delta_2}{3} \cdot \zeta(k_u(w)) \eta(\ell_u(w)) + \frac{2 \delta_2 - 2 \delta_1 \delta_2}{3} \cdot  \zeta(k_u(w)) \eta(\ell_u(w) + 1) \\
& \hspace{8em} - \frac{1}{3} a(k_u(w), \ell_u(w)) \leq 0,
\end{align*}
which can be implied by 
\begin{equation}
a(0, 2) \geq (2 \delta_1 + 2 \delta_2 - 2 \delta_1 \delta_2) \cdot \eta(2) + (2 \delta_2 - 2 \delta_1 \delta_2) \cdot \eta(3)
\end{equation}
because we have ensured that $a(k, \ell)$ is non-decreasing over $k$ and $\ell$.

\subsection{Approximate Dual Feasibility}
The remaining set of constraints for the factor-revealing LP is to satisfy the approximate dual feasibility of Lemma~\ref{lem:primaldual}. Consider an edge $(u, v) \in E$. We want to ensure that, after $v$ gets processed, $\alpha_u + \beta_v \geq \Gamma w_{uv}$ is maintained. Let us fix a timestep after $v$ gets processed.

Let $k_u(w)$ and $\ell_u(w)$ be the number of times that $u$ was inserted into the two-way OCS and the three-way OCS with weight level at least $w$ until $v$ arrives, and let $k'_u(w)$ and $\ell'_u(w)$ be those values at the beginning of the fixed timestep. We know for sure that $k'_u(w) \geq k_u(w)$ and $\ell'_u(w) \geq \ell_u(w)$ for all $w$, but these bounds can be strengthened under specific situations. Recall that, for $k' \geq k$ and $\ell' \geq \ell$, we have $a(k', \ell') \geq a(k, \ell)$ due to monotonicity we will later guarantee.

\subparagraph*{The Boundary Case.} In case of $k'_u(w) = \ell'_u(w) = \infty$, we will guarantee the following condition in order for $\alpha_u$ to cover $\Gamma w_{uv}$:
\[
\alpha_u \geq \int_0^{w_{uv}} a(\infty, \infty) dw \geq \Gamma \cdot w_{uv},
\]
which is equivalent to
\begin{equation} \label{weighted:const41}
a(\infty, \infty) \geq \Gamma.
\end{equation}

\subparagraph*{Case 1. $v$ is matched via the three-way OCS, but $u$ is not inserted.}
In this case, we can use the trivial bounds: $k'_u(w) \geq k_u(w)$ and $\ell'_u(w) \geq \ell_u(w)$ for all $w$. Observe also that, in this case, we have $\beta_v \geq 3 \delRthreeBeta{u}{v}$ by construction of the algorithm. Therefore, to ensure $\alpha_u + \beta_v \geq \Gamma w_{uv}$ at the fixed timepoint, it suffices to have
\[
\int_0^{w_{uv}} a(k_u(w), \ell_u(w)) dw + 3 \int_0^{w_{uv}} b(k_u(w), \ell_u(w)) dw \geq \Gamma \cdot w_{uv}.
\]
We will instead include the following stronger constraints: for every $k$ and $\ell$,
\begin{equation} \label{weighted:const42}
a(k, \ell) + 3 b(k, \ell) \geq \Gamma.
\end{equation}

\subparagraph*{Case 2. $v$ is matched via the three-way OCS, and $u$ is inserted.}
Note that we can strengthen the bounds in this case. Since $u$ was passed to the three-way OCS with weight level $w_{uv}$, we have $\ell'_u(w) \geq \ell_u(w) + 1$ for $w \leq w_{uv}$, instead. (Recall that the fixed timestep is after $v$ gets processed.) Therefore, we have
\[
\alpha_u \geq \int_0^{w_{uv}} a(k_u(w), \ell_u(w) + 1) dw + \int_{w_{uv}}^\infty a(k_u(w), \ell_u(w)) dw.
\]
Note that this case happens when $\beta_v = \betaRthree \geq \betaD \geq \delDBeta{u}{v}$, implying that
\[
\beta_v \geq \delDBeta{u}{v} = \sigD \int_0^{w_{uv}} b(k_u(w), \ell_u(w)) dw - \frac{\sigD}{3} \int_{w_{uv}}^\infty a(k_u(w), \ell_u(w)) dw.
\]
Observe that
\[
\sigD \leq \frac{3 \sigRtwo}{3 - \sigRtwo} = 3 \left( \frac{3}{3 - \sigRtwo} - 1 \right) \leq 3,
\]
where the inequality follows from $0 \leq \sigRtwo \leq 3/2$. Combining these inequalities gives us
\[
\alpha_u + \beta_v \geq \int_0^{w_{uv}} a(k_u(w), \ell_u(w) + 1) dw + \sigD \int_0^{w_{uv}} b(k_u(w), \ell_u(w)) dw.
\]
Thus, we can cover this case by ensuring, for every $k$ and $\ell$,
\begin{equation} \label{weighted:const43}
a(k, \ell + 1) + \sigD \cdot b(k, \ell) \geq \Gamma.
\end{equation}

\subparagraph*{Case 3. $v$ is matched via the two-way OCS, but $u$ is not inserted.}
Since $u$ is not involved when $v$ is processed, we have the trivial bounds. By construction of our algorithm, we can obtain
\[
\beta_v = \delRtwoBeta{\ubar_1}{v} + \delRtwoBeta{\ubar_1}{v} = \sigRtwo \left( \delRthreeBeta{\ubar_1}{v} + \delRthreeBeta{\ubar_2}{v} \right) \geq \delRthreeBeta{\ubar_1}{v} + \delRthreeBeta{\ubar_2}{v} + \delRthreeBeta{u}{v}.
\]
By rearranging the terms, we can derive
\[
3 \delRthreeBeta{u}{v} \leq 3(\sigRtwo - 1) \left( \delRthreeBeta{\ubar_1}{v} + \delRthreeBeta{\ubar_2}{v} \right) \leq \sigRtwo \left( \delRthreeBeta{\ubar_1}{v} + \delRthreeBeta{\ubar_2}{v} \right) = \beta_v,
\]
where the second inequality comes from the fact that $\sigRtwo \leq 3/2$. Note that this case is now subsumed by Case 1.

\subparagraph*{Case 4. $v$ is matched via the two-way OCS, and $u$ is inserted.}
Since $u$ was passed to the two-way OCS with weight level $w_{uv}$, we have a stronger bound: for $w \leq w_{uv}$, $k'_u(w) \geq k_u(w) + 1$. This implies
\[
\alpha_u \geq \int_0^{w_{uv}} a(k_u(w) + 1, \ell_u(w)) dw + \int_{w_{uv}}^\infty a(k_u(w), \ell_u(w)) dw.
\]
We remark that this case happens when $\beta_v = \betaRtwo \geq \betaD \geq \delDBeta{u}{v}$, and thus we can obtain
\[
\beta_v \geq \delDBeta{u}{v} = \sigD \int_0^{w_{uv}} b(k_u(w), \ell_u(w)) dw - \frac{\sigD}{3} \int_{w_{uv}}^\infty a(k_u(w), \ell_u(w)) dw.
\]
As can be seen in Case 2, we know $\sigD \leq 3$, yielding
\[
\alpha_u + \beta_v \geq \int_0^{w_{uv}} a(k_u(w) + 1, \ell_u(w)) dw + \sigD \int_0^{w_{uv}} b(k_u(w), \ell_u(w)) dw,
\]
which can be guaranteed by the following constraints: for every $k$ and $\ell$,
\begin{equation} \label{weighted:const44}
a(k + 1, \ell) + \sigD \cdot b(k, \ell) \geq \Gamma.
\end{equation}

\subparagraph*{Case 5. $v$ is matched deterministically not with $u$.}
Observe that, by construction, we have
\[
\beta_v = \delDBeta{\ustar}{v} = \frac{\sigD}{\sigRtwo} \delRtwoBeta{\ustar}{v} \geq \delRtwoBeta{\ustar}{v} + \delRtwoBeta{u}{v},
\]
implying that
\[
\delRtwoBeta{u}{v} \leq \left( \frac{\sigD}{\sigRtwo} - 1 \right) \delRtwoBeta{\ustar}{v}.
\]
We can thus derive
\[
3 \delRthreeBeta{u}{v} \leq \frac{3}{\sigRtwo} \left( \frac{\sigD}{\sigRtwo} - 1 \right) \delRtwoBeta{\ustar}{v} \leq \frac{\sigD}{\sigRtwo} \delRtwoBeta{\ustar}{v} = \beta_v,
\]
where the second inequality holds because $\sigD \leq 3 \sigRtwo / (3 - \sigRtwo)$. The rest of the argument immediately follows from Case 1.

\subparagraph*{Case 6. $v$ is matched deterministically with $u$.}
Due to the dual update rule, we have
\[
\alpha_u \geq \int_0^{w_{uv}} a(\infty, \infty) dw,
\]
which can be covered by Constraint~\eqref{weighted:const41} of the boundary case.

\subparagraph*{Case 7. $v$ remains exposed.}
In this case, we have $\beta_v = 0$ and $\delDBeta{u}{v} < 0$. Since $\sigD > 0$, we also have $\delRthreeBeta{u}{v} < 0$, implying that $\beta_v \geq 3 \delRthreeBeta{u}{v}$. This case also follows Case 1.

\subsection{A Factor-Revealing LP}
\anhc{Similarly to Appendix~\ref{ss:ufrl}, we write a factor-revealing LP that determines the values of $a$ and $b$. In order to keep the LP finite, we will choose some $\kmax\geq 3$ and $\ellmax\geq 3$ and consider $a(k,\ell)$ and $b(k,\ell)$ only for $k\leq \kmax$ and $\ell\leq\ellmax$. We will define $a(k,\ell):=a(\kmax,\ellmax)$ and $b(k,\ell):=0$ if $k>\kmax$ or $\ell>\ellmax$. We will technically create a few additional variables $a(k,\ell)$ for some $k>\kmax$ and $\ell>\ellmax$ when they appear on constraints; even then their values will be constrained as defined: $a(k,\ell):=a(\kmax,\ellmax)$ and $b(k,\ell):=0$.

\begin{lemma}
The competitive ratio of our algorithm is greater than or equal to the value of the following LP. Let $R(k_0,\ell_0):=\{(k,\ell)\mid k_0\leq k\leq \kmax, \ell_0\leq \ell\leq\ellmax\}$ and $R:=R(0,0)=\{(k,\ell)\mid 0\leq k\leq \kmax, 0\leq \ell\leq\ellmax\}$.

\begin{align}
\text{maximize } & \Gamma  & \nonumber\\
\text{subject to }
& a(k,\ell) = a(\kmax,\ellmax), & \forall (k,\ell)\notin R,\label{wfl37}\\
& b(k,\ell) =0, & \forall (k,\ell)\notin R,\label{wfl38}\\
& a(k,\ell) \leq a(k+1,\ell), & \forall (k,\ell)\in R,\label{wfl39}\\
& a(k,\ell) \leq a(k,\ell+1), & \forall (k,\ell)\in R,\label{wfl40}\\
& a(0, 0)  = 0, & \label{wfl41}\\
 & a(\kmax, \ellmax) - a(k, \ell) + \sigD b(k, \ell)  \leq \zeta(k) \eta(\ell), & \forall (k, \ell) \in R, \label{wfl42}\\
 & a(1, \ell) - a(0, \ell) + \sigRtwo b(0, \ell)  \leq \frac{\eta(\ell)}{2}, & \forall \ell=0,\ldots,\ellmax, \label{wfl43}\\
 & a(k + 1, \ell) - a(k, \ell) + \sigRtwo b(k, \ell)  \leq \frac{1 + \gamma}{2} \zeta(k) \eta(\ell), &\forall (k,\ell)\in R(1,0)\label{wfl44}\\
 & a(1, 0)  \geq \frac{3 \gamma}{4 \sigRtwo}, \label{wfl45}\\
 & a(k, 1) - a(k, 0) + b(k, 0)  \leq \frac{\zeta(k)}{3}, & \forall k =0,\ldots,\kmax, \label{wfl46}\\
 & a(k, 2) - a(k, 1) + b(k, 1)  \leq \frac{2 + 4 \delta_1}{9} \cdot \zeta(k), & \forall k=0,\ldots,\kmax, \label{wfl47}\\
 & a(k, \ell + 1) - a(k, \ell) + b(k, \ell)  & \nonumber \\
 & \hspace{5em} \leq \frac{1 + 2\delta_1 + 2\delta_2 - 2\delta_1 \delta_2}{3} \cdot \zeta(k) \eta(\ell), & \forall (k,\ell)\in R(0, 2), \label{wfl48}\\
 & a(0, 1)  \geq 2 \delta_1 \eta(1) + 2 (\delta_2 - \delta_1 \delta_2) \eta(2), \label{wfl49}\\
 & a(0, 2)  \geq 2(\delta_1 + \delta_2 - \delta_1 \delta_2) \eta(2) + 2(\delta_2 - \delta_1 \delta_2) \eta(3), \label{wfl50}\\
 & a(\kmax, \ellmax)  \geq \Gamma, & \label{wfl51}\\
 & a(k, \ell) + 3b(k, \ell)  \geq \Gamma, & \forall (k, \ell) \in R, \label{wfl52}\\
 & a(k, \ell + 1) + \sigD b(k, \ell)  \geq \Gamma, & \forall (k, \ell) \in R, \label{wfl53}\\
 & a(k + 1, \ell) + \sigD b(k, \ell)  \geq \Gamma, & \forall (k, \ell) \in R, \label{wfl54}\\
 & a(k, \ell)  \geq 0, & \forall (k, \ell) \in R, \label{wfl55}\\
 & b(k, \ell)  \geq 0, & \forall (k, \ell) \in R.\label{wfl56}
\end{align}
\end{lemma}
We create only those variables required by Constraints~\eqref{wfl39}-\eqref{wfl54}. Constraints~\eqref{wfl39} and \eqref{wfl40} impose monotonicity of $a$ over $k$ and over $\ell$. Constraints~\eqref{wfl41}-\eqref{wfl54} are derived from \eqref{weighted:const0}-\eqref{weighted:const44}.
Note that it suffices to impose Constraints~\eqref{wfl37}-\eqref{wfl40}, \eqref{wfl42}-\eqref{wfl44}, and \eqref{wfl46}-\eqref{wfl48} only for $k\leq\kmax$ and $\ell\leq\ellmax$ due to our choice of $a(k,\ell):=a(\kmax,\ellmax)$ and $b(k,\ell):=0$ for $(k,\ell)\notin R$. Constraints~\eqref{wfl52}-\eqref{wfl54} are trivially satisfied for $(k,\ell)\notin R$ because of Constraint~\eqref{wfl51}.}

With $\kmax = 25$ and $\ellmax = 25$, we obtain $\Gamma = 0.50930725$, completing the proof of Theorem~\ref{thm:wwmm}.

\end{document}